\documentclass[usenatbib]{mnras}

\usepackage{newtxtext,newtxmath}

\usepackage[T1]{fontenc}

\DeclareRobustCommand{\VAN}[3]{#2}
\let\VANthebibliography\thebibliography
\def\thebibliography{\DeclareRobustCommand{\VAN}[3]{##3}\VANthebibliography}

\usepackage{graphicx}	        
\usepackage{amsmath}	        
\usepackage{bm}                 
\usepackage{caption}            
\usepackage{subcaption}
\usepackage{color}              
\usepackage[normalem]{ulem}     
\usepackage{nicefrac, xfrac}    

\graphicspath{{Graphs/}}

\title[3D Force-Free Neutron Star Magnetospheres]{Modelling 3D Force-Free Neutron Star Magnetospheres}

\author[P. Stefanou et al.]{
Petros Stefanou,$^{1}$\thanks{E-mail: petros.stefanou@uv.es}
Jose A. Pons,$^{2}$
Pablo Cerdá-Durán$^{1}$
\\
$^{1}$Departament d'Astronomia i Astrofísica, Universitat de València, Dr. Moliner 50, E-46100, Burjassot, València, Spain\\
$^{2}$Departament de Física Aplicada, Universitat d'Alacant, Ap. Correus 99, E-03080 Alacant, Spain\\
}

\date{Accepted XXX. Received YYY; in original form ZZZ}

\pubyear{2021}

\begin{document}
\label{firstpage}
\pagerange{\pageref{firstpage}--\pageref{lastpage}}
\maketitle

\begin{abstract}

Magnetars exhibit a variety of transient high-energy phenomena in the form of bursts, outbursts, and giant flares. 
It is a common belief that these events originate in the sudden release of magnetic energy due to the rearrangement of a twisted magnetic field. 
We present global models of a 3D Force-Free (FF) non-linear twisted magnetar magnetosphere. 
We solve the FF equations following the Grad-Rubin approach in a compactified spherical domain. 
Appropriate boundary conditions are imposed at the surface of the star for the current distribution and the magnetic field. 
Our implementation is tested by reproducing various known analytical as well as axisymmetric numerical results. 
We then proceed to study general 3D models with non-axisymmetric current distributions, such as fields with localised twists that resemble hotspots at the surface of the star, and we examine characteristic quantities such as energy, helicity and twist. Finally, we discuss implications on the available energy budget, the surface temperature and the diffusion timescale, which can be associated with observations.

\end{abstract}




\section{Introduction}\label{sec:introduction}

Magnetars are a special class of isolated neutron stars characterised by their strong magnetic field, typically of the order of $10^{14}$ G or more. 
They are relatively young and slowly rotating with periods of the order of $1-10$ s. 
This combination of characteristics suggests a very rapid spin-down of the star due to intense magnetic braking, placing these objects in the upper right corner of the $P\Dot{P}$ diagram. 
Contrary to other classes of neutron stars, where rotation or accretion provides the necessary energy reservoir, the powering source of magnetar emission is the dissipation of their enormous magnetic field \citep{Thompson_Duncan_1995, Thompson_Duncan_1996}. 
Magnetars are observed mainly in the X-ray and soft $\gamma$-ray bands and show a plethora of transient phenomena, highly variable and with unpredictable occurrence \citep{Mereghetti_Pons_Melatos_2015, Kaspi_Beloborodov_2017}. 
This transient activity is believed to be linked to the presence of a twisted, current-filled magnetosphere supported by the long-term evolution of the interior magnetic field \citep{Beloborodov_2007}. 
In particular, the evolution of the magnetic field in the neutron star crust gradually displaces the footprints of the magnetic field lines, feeding the magnetosphere with energy and introducing currents in the twisted region. 
At some critical point, a large-scale reconfiguration of the magnetic field takes place, releasing the stored energy in the form of high-energy radiation. 
Whether the triggering mechanism of these events is originated in the star's crust, caused by a mechanical failure due to the build-up of magnetic stresses \citep{Perna_Pons_2011}, or is magnetospherically-driven, where the large magnetospheric twist becomes unsustainable and reconnection takes place \citep{Lyutikov_2003}, is still a topic of investigation. 

Nevertheless, an important task in revealing the physics of the magnetar bursting activity is the study of equilibrium solutions of the twisted magnetospheric field for a given field and current distribution at the surface of the star. 
Various works in the literature have attacked this problem in axial symmetry, by solving a Grad-Shafranov equation. 
\citet{Glampedakis_Lander_Andersson_2014} provided stationary axisymmetric solutions extending from the interior of the star to the magnetosphere. A similar effort was done by
\citet{Pili_Bucciantini_Del_Zanna_2015}, but in a GRMHD framework.
\citet{Kojima_2017} studied magnetospheric solutions considering effects of general relativity.
\citet{Akgun_Miralles_Pons_Cerda_2016, Akgun_Cerda_Miralles_Pons_2017} provided a detailed parametric study, varying the extent of the current-filled region, the non-linearity of the model and the strength of the toroidal field. They showed that the magnetosphere could support twists up to $\sim 1.5$ rad and could reach total energies up to $\sim 25\%$ more than the vacuum solution. A second interesting finding was that the Grad-Shafranov equation is degenerate and solutions with different total energies could exist for the same model parameters. By performing dynamical simulations on these models, \citet{Mahlmann_Akgun_Pons_Aloy_Cerda_2019} confirmed that only the lowest energy branch of these solutions is stable. Similar instabilities have been observed in a dynamically twisted magnetospheres for values of $\sim 1$~rad \citep{Parfrey2012,Parfrey2013,Carrasco2019}. 

In all these works, the general consensus is that beyond a maximum twist, no magnetospheric solution is achievable. This is a strong sign of the existence of some critical point where solutions may become unstable, leading to a sudden, large scale reorganisation of the magnetic field.

Some observational features suggest that the topology of the twisted magnetosphere is highly unlikely to be axisymmetric \citep{Tiengo_et_al_2013} and rather resembles the sun's coronal configuration, with localised, closed magnetic loops \citep{Younes_et_al_2022}. Theoretical results on the evolution of the internal magnetic field also lean towards the same conclusion. For example, \citet{Lander_Gourgouliatos_2019} argue that the magneto-plastic evolution in the NS crust could produce magnetospheric twist that are local and non-axisymmetric.

The evolution of the magnetic field inside the star plays a key role to the eventual magnetospheric configuration. The deviation of the exterior magnetic field from the standard, current-free, dipole picture depends on the current distribution and the toroidal field at the surface, which, in turn, are a result of the magneto-thermal evolution of the magnetic field in the crust, under the effect of Hall drift and Ohmic dissipation \citep{Vigano_Rea_Pons_et_al_2013}. Additionally, allowing currents to flow through the crust (i.e using a non-vacuum solution as a surface boundary condition for the magneto-thermal evolution) can significantly affect the internal magnetic field. Sadly, the vastly different physical conditions in the two regions, mainly the plasma density and the Alfvén crossing time, make it difficult to perform joint studies of the evolution of the magnetic field both inside and outside the star. As a possible solution, \cite{Akgun_Cerda_Miralles_Pons_2018a,Akgun_Cerda_Miralles_Pons_2018b} performed 2D simulations of the  magneto-thermal evolution of the NS crust coupled to a FF magnetospheric solution of the Grad-Shafranov equation. In this work, we aim at giving a step forward in our understanding of magnetar magnetospheres by presenting the first 3D models in this context.

The following sections are organised as follows. In section \ref{sec:equations} we state the problem and we give an overview of the mathematical equations that describe it. In section \ref{sec:method} we present the numerical method that we employ and go over the details of the numerical setup. We perform tests for the reliability of our numerical implementation in section \ref{sec:tests}. Sections \ref{sec:results} and \ref{sec:discussion} are dedicated to our results and an analysis of their most interesting features. We conclude with some final remarks and future suggestions in section \ref{sec:conlcusion}.



\section{Basic Equations}\label{sec:equations}

\subsection{Force-free field}

The magnetospheres of isolated NSs are generally considered to be dominated by the magnetic field. The physical conditions in the magnetosphere, i.e. low mass of the charge carriers, low density of the magnetospheric plasma and non-relativistic co-rotational speed, allow us to neglect gravity, thermal pressure and particle inertia as they are several orders of magnitude smaller than the \textbf{electromagnetic} forces. In addition, in the particular case of magnetars, due to their slow rotation, the light cylinder is expected to lie far away from the star’s surface and rotationally induced electric fields are not expected to play any significant role to the dynamics of the magnetosphere. With these assumptions, the equation that describes the force balance in the magnetosphere is reduced to simply demanding the magnetic force to be equal to zero
\begin{equation}
    \bm{J} \times \bm{B} = 0 
    \label{eq:laplace},
\end{equation}
where
\begin{equation}
    \bm{J} = \frac{c}{4 \pi} \nabla \times \bm{B}
    \label{eq:current}
\end{equation}
is the electric current. In addition, the magnetic field needs to be divergence-free, as described by the solenoidal condition
\begin{equation}
    \nabla \cdot \bm{B} = 0
    \label{eq:solenoidal}.
\end{equation}

Eq. \eqref{eq:laplace} merely implies that the current is always parallel to the magnetic field
\begin{equation}
    \nabla \times \bm{B} = \alpha(\bm{r}) \bm{B}
    \label{eq:force_free},
\end{equation}
where $\alpha$ is a parameter, usually called the FF parameter or the current parameter, associated with the strength of the twist (or equivalently, the toroidal component of the magnetic field in axisymmetry) in the magnetosphere. Eq. \eqref{eq:force_free} is the so-called FF equation that describes the magnetic field of a twisted magnetosphere in the FF regime.
It is straightforward to show, from Eq.s \eqref{eq:solenoidal} and \eqref{eq:force_free}, that $\alpha$ has to obey the constraint
\begin{equation}
    \bm{B} \cdot \nabla \alpha = 0
    \label{eq:alpha_constraint},
\end{equation}
which states that $\alpha$ is constant along any particular magnetic field line. In other words, the field lines lie on isocontours of the parameter $\alpha$. 
We note that Eq. \eqref{eq:alpha_constraint} introduces an implicit non-linearity to the problem through the dependence of $\alpha$ on the magnetic field. In general, three distinct cases of Eq. \eqref{eq:force_free} can be studied:

\begin{enumerate}
    \item $\alpha = 0$:  Current-free field ($\bm{J}=0$). \label{it:alpha_zero}
    
    The equation reduces to $\nabla \times \bm{B} = 0$, which describes a potential field consisting of a linear combination of multipoles.
    
    \item $\alpha =$ const. :  Linear FF field. \label{it:alpha_const}
    
    The equation reduces to the Helmholtz equation $\nabla^2\bm{B} + a^2\bm{B} = 0$, which accepts solutions in terms of the spherical Bessel functions.
    
    \item $\alpha = \alpha(\bm{r})$: Non-linear FF field.\label{it:alpha_general}
    
    The equation is non-linear. One has to make use of numerical methods in order to determine the solution.
\end{enumerate}

The first two cases, while useful to give insight, do not correspond to magnetospheres that can appear in nature. Case \ref{it:alpha_zero} describes a current-free, vacuum magnetosphere which is not consistent with the well established idea of the existence of a twist and cannot explain the magnetospheric emission observed from magnetars. Case \ref{it:alpha_const} results in solutions that do not decay with distance but are oscillatory at infinity and therefore, conservation of energy cannot be enforced.

Case \ref{it:alpha_general} is the only one that is physically relevant and can describe a real magnetar magnetosphere. Solving this non-linear equation, especially in a 3D,  non-axisymmetric setup, is a challenging task. We describe the numerical method with which we attack this problem in section \ref{sec:method}.

\subsection{Compactification.}\label{sec:geometry}

The absence of elements such as the light cylinder, an accretion
disk, etc., highly favours the use of spherical coordinates for the
description of the problem. However, since the physical solutions need to decay sufficiently fast at large distances, one has to place the outer boundary far away to make sure that the fields satisfy the required asymptotic solution. This usually comes at the cost of lowering the 
resolution, since achieving both is computationally expensive. In
addition, special attention must be paid to the implementation of boundary conditions to avoid contamination of the solution from inaccuracies in the boundaries. These problems can severely hinder any
attempt to achieve accurate solutions, especially in 3D, where the computational cost is an important factor.

An effective strategy is to compactify the radial coordinate by mapping the entire physical domain, from the stellar surface $R$ up to infinity, to a sphere of radius $R$, with the origin as its center. This is achieved by the following transformation:
\begin{equation}
    r \in [R, \infty] \longrightarrow q \in [0, R],
\end{equation}
which can be realised by the, rather simple, transformation formula
\begin{equation}
    q = \frac{R^2}{r},
\end{equation}
while the angular coordinates $\theta$ and $\varphi$ remain unchanged

One can easily check that in both systems the surface is given by the same value, $r = R \Leftrightarrow q = R$ and that infinity in the normal radial coordinate is represented by just one point, the origin, in the compact radial coordinate, $r = \infty \Leftrightarrow q = 0$.

By introducing this coordinate system and writing the equations in terms of $q$ instead of $r$ we alleviate the difficulties of imposing boundary conditions at large distances. The price to pay is small as only minor changes are introduced in the equations (see next section). With this approach, all the quantities that need to be specified at infinity and are expected to decay with radius can be simply put to zero at $q = 0$.

An additional advantage of this coordinate transformation is that quantities that are inversely proportional to some power of the radial coordinate $\propto r^{-p}$ now become polynomials in terms of the coordinate $q$. This makes finite differences approximations of their derivatives significantly more accurate. 

Hereafter, we use compactified spherical coordinates as our coordinate system unless stated otherwise. 
In order to maintain a clear and simple notation, we assume that distances are measured in terms of the stellar surface so that $R = 1$. Therefore $q \in [0,1]$ and the surface of the star corresponds to $q = 1$.

\section{Method}\label{sec:method}

There exists a number of different numerical methods in the literature that can be used for solving equations similar to Eq. \eqref{eq:force_free}, which are mainly encountered in the area of solar physics for the reconstruction of coronal magnetic fields \citep[see][for a detailed review on this topic]{Wiegelmann_Sakurai_2021}. For our purposes, a suitable method is the so-called \textit{Grad-Rubin} method.

\subsection{General Idea}

The Grad-Rubin method was originally proposed by \citet{Grad_Rubin_1958} in the context of laboratory plasma physics but implemented for the first time by \citet{Sakurai_1981} and later by \citet{Amari_Boulmezaoud_Mikic_1999} to determine local magnetic field configurations in the solar corona. 

The approach that we follow in this work is largely based on the one described in \citet{Amari_Boulmezaoud_Mikic_1999, Amari_Boulmezaoud_Aly_2006, Amari_Aly_Canou_Mikic_2013} for the reconstruction of magnetic fields in the solar corona. In this approach, the magnetic field $\bm{B}$ is described in terms of a vector potential $\bm{A}$, given by
\begin{align}
    \bm{B} &= \nabla \times \bm{A}\label{eq:vector_potential}, 
\end{align} 
We choose a convenient gauge condition corresponding to the Coulomb gauge:
\begin{align}
    \nabla \cdot \bm{A} &= 0, & \textrm{($q\le 1$)}\label{eq:gauge} \\
    \nabla_{ang} \cdot \bm{A}_{ang} \Big\vert_{S} &= 0, & (\textrm{surface, $q=1$})\label{eq:gauge_surface},
\end{align}
where the subscript $S$ denotes the surface of the NS at $q = 1$ and 
\begin{equation}
    \nabla_{ang} \cdot \bm{A}_{ang} = \frac{q}{\sin{\theta}} \frac{\partial}{\partial \theta} (A_\theta \sin{\theta}) + \frac{q}{\sin{\theta}} \frac{\partial A_\phi}{\partial \theta}
\end{equation}
is the part of the divergence that involves only the angular derivatives. 

Boundary conditions need to be provided at the surface of the NS. These are: a) $b_S(\theta,\phi$), the radial component of the magnetic field and b) $\alpha_S (\theta, \phi)$, the value of $\alpha$ at the surface. 
Condition (b), in particular, determines the region where currents flow, and
must be implemented in a way that ensures that the two footprints of any given magnetic field line have the same value of $\alpha$.

As argued in \citet{Bineau_1972}, the existence of a solution is guaranteed for such boundary conditions if, in addition, the values of $\alpha$ are sufficiently small and the non-linearity is relatively weak. 

In the numerical procedure devised by Grad and Rubin, the system of equations \eqref{eq:force_free} and \eqref{eq:alpha_constraint} is solved by means of a fixed-point 
iteration, in which one solves alternatively Eq.~\eqref{eq:alpha_constraint} for $\alpha$ with $\bm{B}$ fixed (hyperbolic part) and Eq.~\eqref{eq:force_free} for $\bm{B}$ with $\alpha$ fixed (elliptic part), until convergence to a global solution is achieved. We described next the numerical details of both parts separately.

As discussed in section \ref{sec:geometry}, our implementation uses compactified spherical coordinates $(q, \theta, \varphi$). The range of our domain is $q \in [0,1], \ \theta \in [0,\pi],  \ \varphi \in [0, 2 \pi]$, where we discretize all our equations in a grid $\{q_i, \theta_j, \phi_k\}$, with $i=1, ..., m$, $j=1 ,..., n$ and $k=1, ..., o$. The surface and infinity are the only real boundaries on our domain, where physical boundary conditions need to be imposed. The boundaries on $\theta$ and $\varphi$ are just internal boundaries, where care needs to be taken to ensure that the solution crosses them smoothly and there are no singularities that may cause divergence of the solution at the axis.

\subsection{Hyperbolic Part}

For a given magnetic field $\bm{B}$, Eq. \eqref{eq:alpha_constraint} implies that, whatever the solution of the FF equation, it has to fulfil that $\alpha$ is constant along magnetic field lines. In particular, at the two footprints of each magnetic field line the value of $\alpha$ has to be identical. This sets a constraint on the possible values of $\alpha$ at the surface (the boundary condition $\alpha_0$). The problem is that this constraint cannot be set a priory, because it needs the knowledge of the magnetic field line structure, which is part of the solution that we want to obtain. Setting an arbitrary function $\alpha_0$ could very easily lead to boundary conditions with no FF magnetospheric solution possible.

For this reason, $\alpha_S$ must be provided only at a region of a given magnetic polarity, i.e. either the region with $B_q>0$ (positive polarity) or the region with $B_q<0$ (negative polarity). To simplify the arguments, we always assume the former, since the other case is analogous.

In this case, an initial value problem for $\alpha$ can be written in the form
\begin{align}
    \bm{B}^{(n)} \cdot \nabla \alpha^{(n+1)} &= 0, & \textrm{(exterior, $q<1$)} \label{eq:hyperbolic}\\
    \alpha^{(n+1)} \Big\vert_{S^+} &= \alpha_S, & \textrm{(surface, $q=1$)}\label{eq:hyperbolic_bc},
\end{align}
where the exponent $n$ denotes the fixed-point iteration number and the subscript $S^+$ denotes the part of the surface where $B_q > 0$.

For our purposes, the most suitable method for solving Eq. \eqref{eq:hyperbolic} is the \textit{Characteristics} or \textit{Field Line Tracing} method, because the characteristic curves of $\alpha$ are non other than the magnetic field lines. 

Every grid point in the domain is threaded by a field line. If the only source of currents is the neutron star surface, then any field line can be traced back to the surface of the star where the values of $\alpha$ are known. Therefore, it is possible to build a numerical method that traces magnetic field lines from any grid point to the surface of the star and assigns the value of $\alpha$ there to the grid point. However, this procedure would be computationally expensive in general. Instead, we have devised a numerical method that is essentially equivalent but significantly more efficient computationally.

The magnetic field lines can be computed by integrating the equation
\begin{equation}\label{eq:field_line_integration}
    \frac{d \bm{x}}{d \lambda} = \frac{\bm{B}}{||\bm{B}||},
\end{equation}
where $\lambda$ is a parametric variable corresponding to the arc length along the magnetic field line. Numerically it is convenient to express and solve this equation in Cartesian coordinates to avoid the coordinate singularities at the axis associated to the compactified coordinates chosen in this work. Details on the transformations from spherical to Cartesian coordinates and vice-versa are given in Appendix~\ref{app:tilt_vector_field}.   

The first part in the procedure is to compute the value of $\alpha$ at all grid points at the surface at $q_1$, a sphere right above the surface of the star. For this purpose we integrate the magnetic field lines starting at every grid point with coordinates $(q_1, \theta_j, \phi_k)$ in the direction opposite to $\bm{B}$. This integration is continued all along the magnetosphere until the surface is reached. Because we are following the field lines backwards, the landing point (footprint) necessarily has $B_q>0$ and hence is part of $S_+$, where the values of $\alpha$ are known. In general the footpoint will not coincide precisely with a grid point, but is is possible to compute it as an interpolation of $\alpha_0$ at the surface. As a result, we get the value of $\alpha$ at all points in $q_1$.

The second part of the procedure builds the solution in the rest of the magnetosphere ($i>1$) computing it layer by layer in the radial direction. To compute $\alpha$ in a radial layer at $q_{i+1}$, from the values of $\alpha$ at $q_i$, we integrate a magnetic field line starting at each point $(q_{i+1}, \theta_j, \phi_k)$ in the direction towards the surface at $q_i$. This integration is very fast in general, because the distance between both surfaces is small. The landing point at $q_i$ will not coincide in general with a grid point, but the value of $\alpha$ can be interpolated within the surface, and assigned to corresponding point at $q_{i+1}$. The procedure is continued until the solution at all grid points is known.

All the integrations (Eq. \eqref{eq:field_line_integration}) are performed using the Adams predictor-corrector methods. All the necessary interpolations are preformed using B-splines of order $k=2$.

\subsection{Elliptic Part}\label{sec:elliptic}


Once $\alpha^{(n+1)}$ is known everywhere, we proceed to compute the value of $\bm{B}^{(n+1)}$ or equivalently $\bm{A}^{(n+1)}$.
The magnetic field can be calculated everywhere in the domain by solving the elliptic boundary value problem 
\begin{align} \label{eq:elliptic_part}
    \nabla \times \bm{B}^{(n+1)} &= \alpha^{(n+1)} \bm{B}^{(n)} \\
    B_q (\theta,\varphi) \Big\vert_{S} &= b_S (\theta, \varphi),\label{eq:boundary_Bq_surface} \\
    \bm{B} \Big\vert_{\infty} &= 0\label{eq:boundary_B_infinity}.
\end{align}
with a fixed source term $\alpha^{(n+1)} \bm{B}^{(n)}$.

We need to express this problem in terms of the vector potential in the selected gauge. Using equations \eqref{eq:vector_potential} and \eqref{eq:gauge} and some vector calculus identities, we can restate Eq. \eqref{eq:elliptic_part} as an equation for the vector potential
\begin{align}
    \nabla^2 \bm{A}^{(n+1)} &= -\alpha^{(n+1)} \bm{B}^{(n)} \label{eq:elliptic} 
\end{align}

Besides, by introducing a poloidal scalar field $\Phi$, which, at the surface, satisfies the condition
\begin{equation}
    \nabla^2_{ang} \Phi \Big\vert_{S} = -b_0 (\theta, \varphi) \label{eq:chi},
\end{equation}
we can calculate the angular components of the vector potential at the surface as follows
\begin{align}
    A_\theta \Big\vert_{S} &= \frac{1}{\sin{\theta}} \frac{\partial \Phi}{\partial \phi}\\
    A_\phi \Big\vert_{S} &= - \frac{\partial \Phi}{\partial \theta}.
\end{align}
More details about these derivations can be found in Appendix \ref{app:poloidal_toroidal_decomposition} (see, in particular, equations \eqref{eq:toroidal_vector_potential} and \eqref{eq:radial_magnetic_field}).
The radial component of the vector potential, from equations \eqref{eq:gauge} and \eqref{eq:gauge_surface}, needs to satisfy the Robin-type boundary condition
\begin{equation}
    \left(2 q A_q - q^2\frac{\partial A_q}{\partial q} \right) \Big\vert_{S} = 0.
\end{equation}
At infinity, all components need to smoothly decay, so the boundary condition \eqref{eq:boundary_B_infinity} is simply replaced by
\begin{equation}
    \bm{A} \Big\vert_{\infty} = 0
\end{equation}

Eq. \eqref{eq:elliptic} is a vector \textit{Poisson} equation with a linear source term, the known current from the previous iteration step.
The three components of \eqref{eq:elliptic} form a system of equations which is coupled, because of terms introduced by the action of the \textit{Laplacian} on a vector in curvilinear coordinates. This is an inconvenience, but our problem was non-linear to begin with and we already had to deal with an iteration procedure for its solution. We can simply pass any additional, coupled terms to the right hand side and solve a scalar \textit{Poisson} equation for each component with a generalised source term. Expanding Eq. \eqref{eq:elliptic}, we get a system of three equations:
\begin{equation}
    \begin{split}
        &\nabla^2 A_q^{(n+1)} - 2 q^2 A_q^{(n+1)}  \\
        = \ &-\alpha^{(n+1)} B_q^{(n)} + 2 q^2 \cot{\theta} A^{(n)}_\theta + 2 q^2 \frac{\partial A^{(n)}_\theta}{\partial \theta} \\
        &+ \frac{2 q^2}{\sin{\theta}}\frac{\partial A^{(n)}_\varphi}{\partial \varphi}\label{eq:vector_laplacian_q},
    \end{split}
\end{equation}

\begin{equation}
    \begin{split}
        &\nabla^2 A_\theta^{(n+1)} - q^2 \frac{A_\theta^{(n+1)}}{\sin^2{\theta}} \\ 
         = \ &-\alpha^{(n+1)} B_\theta^{(n)} - 2 q^2 \frac{\partial A_q^{(n)}}{\partial \theta} + 2 q^2 \frac{\cos{\theta}}{\sin^2{\theta}} \frac{\partial A_\varphi^{(n)}}{\partial \varphi}\label{eq:vector_laplacian_theta},
    \end{split}
\end{equation}

\begin{equation}
    \begin{split}
        &\nabla^2 A_\varphi^{(n+1)} - q^2 \frac{A_\varphi^{(n+1)}}{\sin^2{\theta}} \\ 
         = \ &-\alpha^{(n+1)} B_\varphi^{(n)} -  \frac{2 q^2}{\sin{\theta}}\frac{\partial A_q^{(n)}}{\partial \varphi} - 2 q^2 \frac{\cos{\theta}}{\sin^2{\theta}} \frac{\partial A_\theta^{(n)}}{\partial \varphi}\label{eq:vector_laplacian_phi}.
    \end{split}
\end{equation}
Here, the first term (the current) in the right-hand side of each equation is the actual physical source term. All the other terms in the right-hand side of the equations are by-products of the vector \textit{Laplacian}. The second term in the left-hand side is also a by-product of the vector \textit{Laplacian}, but we chose to keep it there for numerical implementation purposes. 

All elliptic operators are solved using a Scheduled Relaxation Jacobi method \citep{Adsuara2016}. This iterative method has the simplicity of the Jacobi method but with convergence properties similar to the optimal Successive Over-Relaxation (SOR) method. It also has the advantage with respect to the SOR method of not having to optimize any over-relaxation parameter.

\subsection{Divergence Cleaning}

To make sure that the gauge chosen in Eq. \eqref{eq:gauge} is maintained through the iteration process and is not affected by numerical contamination, we introduce a divergence cleaning step after the calculation of the vector potential $\bm{A}$.

If, due to approximation and round-off errors, the divergence of $\bm{A}$ is different from zero, we redefine $\bm{A}$ by introducing a scalar field $\xi$ as follows:
\begin{equation}
    \bm{A}' = \bm{A} - \nabla \xi,
\end{equation}
and we demand that the divergence of the new vector potential $\bm{A}'$ satisfies the gauge $\nabla \cdot \bm{A}' = 0$.
This is only true if 
\begin{equation}
    \nabla^2 \xi = \nabla \cdot \bm{A},
\end{equation}
that is, $\xi$ is the solution of a Poisson equation with a source term equal to the old, non-zero divergence of the vector potential.

With this process, any numerically induced deviation from the selected gauge is compensated by subtracting the parts that contribute to it from the vector potential.

\section{Tests} \label{sec:tests}

Before proceeding to study astrophysically relevant models, we put our code to the test against known solutions. The following test cases will demonstrate that the code is well-behaved and reliable in capturing all the features of a FF non-axisymmetric magnetosphere in compactified spherical coordinates.

\subsection{Test 1: Tilted Dipole}\label{sec:test_tilted_dipole}

The first test aims at providing a benchmark for the code when solving for 3-dimensional magnetic fields. 
In particular, a dipole magnetic field with its axis tilted by an angle $\beta$ around the x-axis. No currents are introduced, so $\alpha = 0$ everywhere for this test. 
We focus on validating other aspects of the code, such as the implementation of the compactified coordinate system, the behaviour of the code close to pathological regions (axis, infinity) and the order of convergence. 
In addition, we calculate the energy and the helicity of the magnetic field to get an estimate of the numerical dissipation introduced.

Tilting a vector field $\bm B46010 (\bm r)$ is a two-step procedure. First a  rotation of the coordinates $(q, \theta, \phi)$ by an angle $-\beta$ is performed, followed by a rotation of the components of the vector $(v_q, v_\theta, v_\phi)$ by an angle $\beta$. The detailed process is described in Appendix \ref{app:tilt_vector_field}. 

\begin{figure}
    \centering
    \includegraphics[width=\columnwidth]{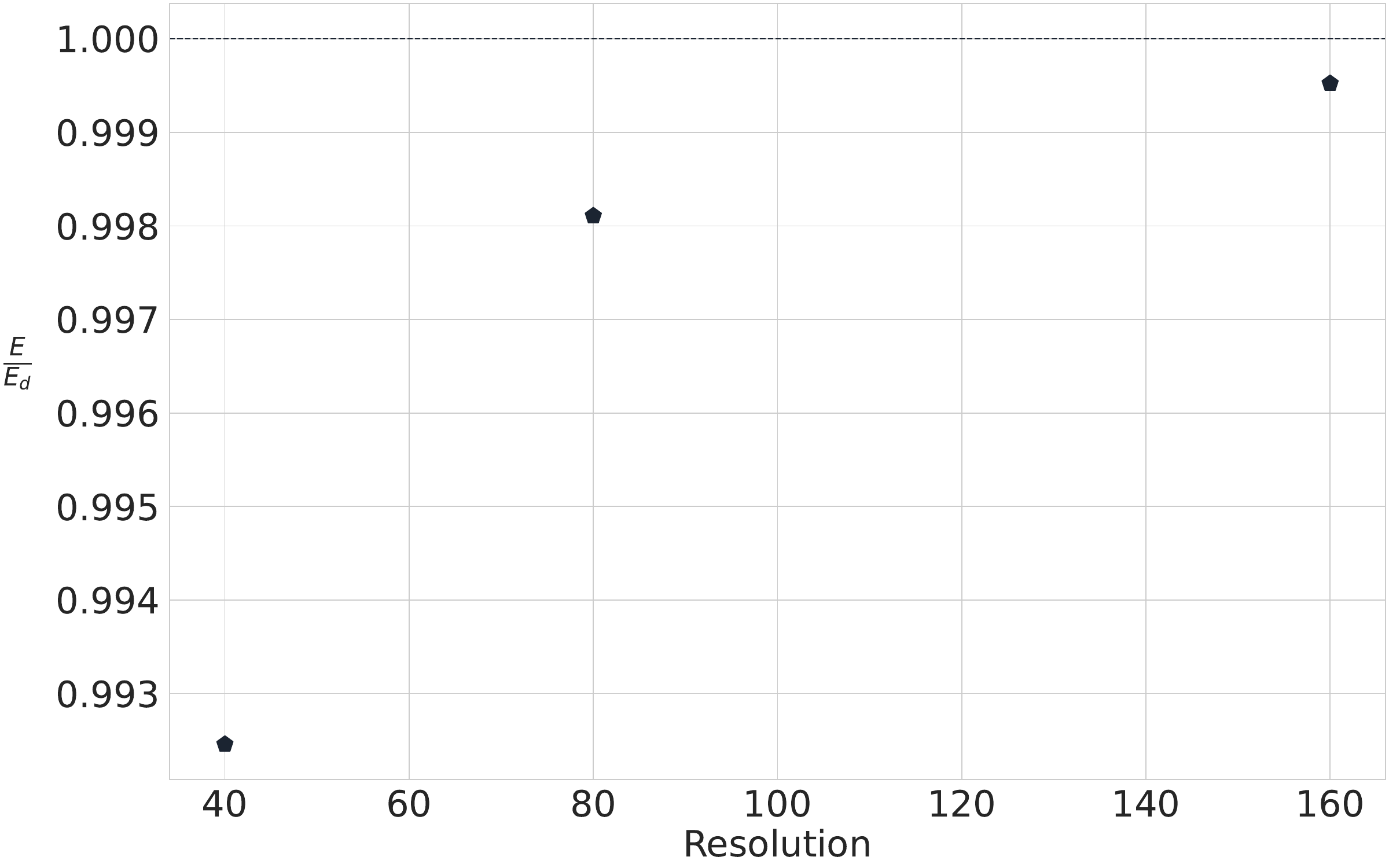}
    \caption{Normalised energy of the tilted dipole solution as a function of the resolution. The dotted line the denotes the analytical solution. The labels of the x-axis show only the radial resolution for clarity.}
    \label{fig:convergence_dipole_tilted_energy}
\end{figure}

\begin{figure}
    \centering
    \includegraphics[width=\columnwidth]{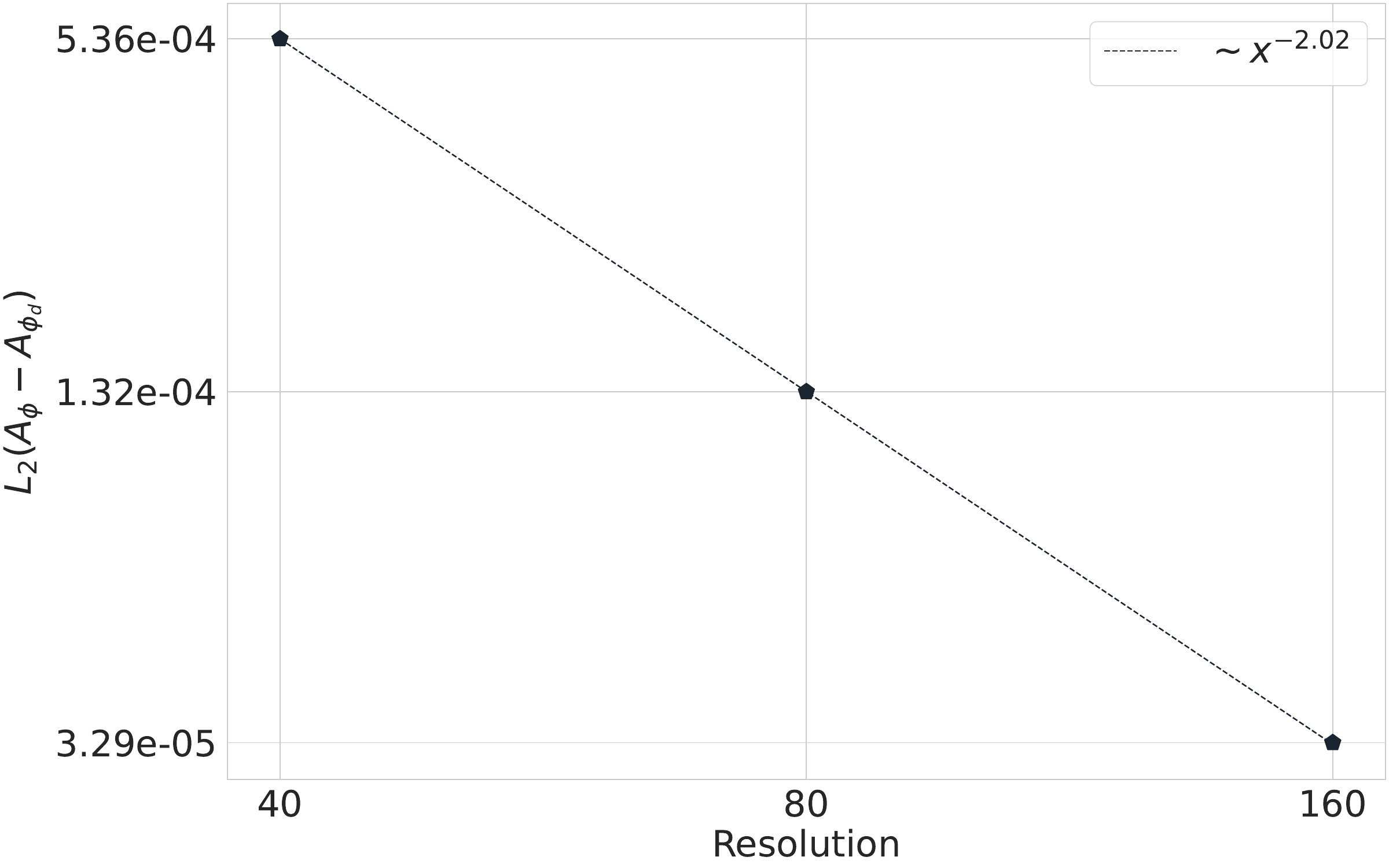}
    \caption{Convergence of the $L_2$ norm of the difference between numerical and analytical solution with resolution. Only the $A_\phi$ component is plotted. Both axes are in logarithmic scale. The dotted line corresponds to a power law fit with index $p = 2.02$, which indicates that the convergence is of second order, as expected by our discretisation.}
    \label{fig:convergence_dipole_tilted_L2}
\end{figure}

We calculate three different solutions for the tilted dipole, with different resolutions: $40 \times 20 \times 40$, $80 \times 20 \times 80$ and $160 \times 80 \times 160$.
Fig. \ref{fig:convergence_dipole_tilted_energy} shows the energy stored in the solution as a function of the resolution. 
It is calculated numerically by the integral
\begin{equation}\label{eq:energy}
    E = \int_\mathcal{V} \frac{B^2}{8 \pi} dV
\end{equation}
over the entire volume of our domain and normalised to the analytical value for a purely dipolar solution
\begin{equation}\label{eq:dipole_energy}
    E_d = \frac{B_0^2 R^3}{3},
\end{equation}
where $B_0$ denotes the surface magnetic field at the equator and $R$ is the radius of the star.
The energy approaches nicely the expected value with increasing resolution. For the case with the highest resolution considered here, the difference between numerical and analytical value is less than 0.1\%. 
We define the helicity of the magnetic field as
\begin{equation}
    H = \int \bm{A} \cdot \bm{B} \ dV.
\end{equation}
Helicity is of the order $\sim 10^{-15}$ for all resolutions and, therefore, very close to the theoretical value of $H_d = 0$.
Fig. \ref{fig:convergence_dipole_tilted_L2} shows the $L_2$ norm of the difference between the numerical and analytical solution. We only plot the $A_\phi$ component of the vector potential, since the other two components are zero and the comparison is not really informative. The convergence of the $L_2$ norm is of second order, as expected. 

All the above metrics indicate that our numerical code is trustworthy and consistent when dealing with current-free solutions. In the next sections we examine cases with $\alpha \neq 0$.

\subsection{Test 2: Axisymmetric FF  Field}\label{sec:test_ff2d}

Twisted magnetospheres in axial symmetry have been studied in detail in a number of works in the past. This test aims at reproducing some of the well-established results obtained in \citet{Akgun_Cerda_Miralles_Pons_2017}, using the Grad-Rubin method.

The two numerical schemes treat the FF problem quite distinctly. In \citet{Akgun_Cerda_Miralles_Pons_2017}, the variable of the problem is a poloidal function $P$ and the current is introduced through a toroidal function $T (P)$, which is determined by three parameters $P_c, \sigma$ and $s$. These parameters control the extent of the current-filled region, the non-linearity of the model and the strength of the current. Under the assumption of axisymmetry and the decomposition of the magnetic field into the scalars $P$ and $T(P)$, Eq. \eqref{eq:force_free} is reduced to the Grad-Shafranov equation for $P$. The vector potential $\bm A$ and the magnetic field $\bm B$ are then reconstructed from $P$ and $T(P)$. In the Grad-Rubin method, Eq. \eqref{eq:force_free} is solved as it is, and the variables of the problem are the vector potential $\bm A$, as well as the current parameter $\alpha$. In the nomenclature of \citet{Akgun_Cerda_Miralles_Pons_2017}, $\alpha$ is the same as $T'(P)$, where the prime denotes the derivative with respect to $P$ (see also equation (8) in \citet{Akgun_Miralles_Pons_Cerda_2016}) and $P$ is related to the azimuthial component of the vector potential through the relation $P = \frac{\sin{\theta}}{q} A_\phi$. It is straightforward to reproduce any of the models studied in \citet{Akgun_Cerda_Miralles_Pons_2017} using our own implementation, by imposing at the surface the current distribution $\alpha = T' (P;P_c, \sigma, s)$ that corresponds to the parameters of that model. 
We remind that $\alpha$ has units of inverse length (we choose to express it in units of the star radius [$R^{-1}$]).
For the purposes of this test, we choose the solution obtained for $P_c = 0.3405074, \sigma = 2$ and $s = 2$ and belongs to the lower energy branch. We will refer to this solution as the FF2D solution from now on. 

\begin{figure}
    \centering
    \includegraphics[width=\columnwidth]{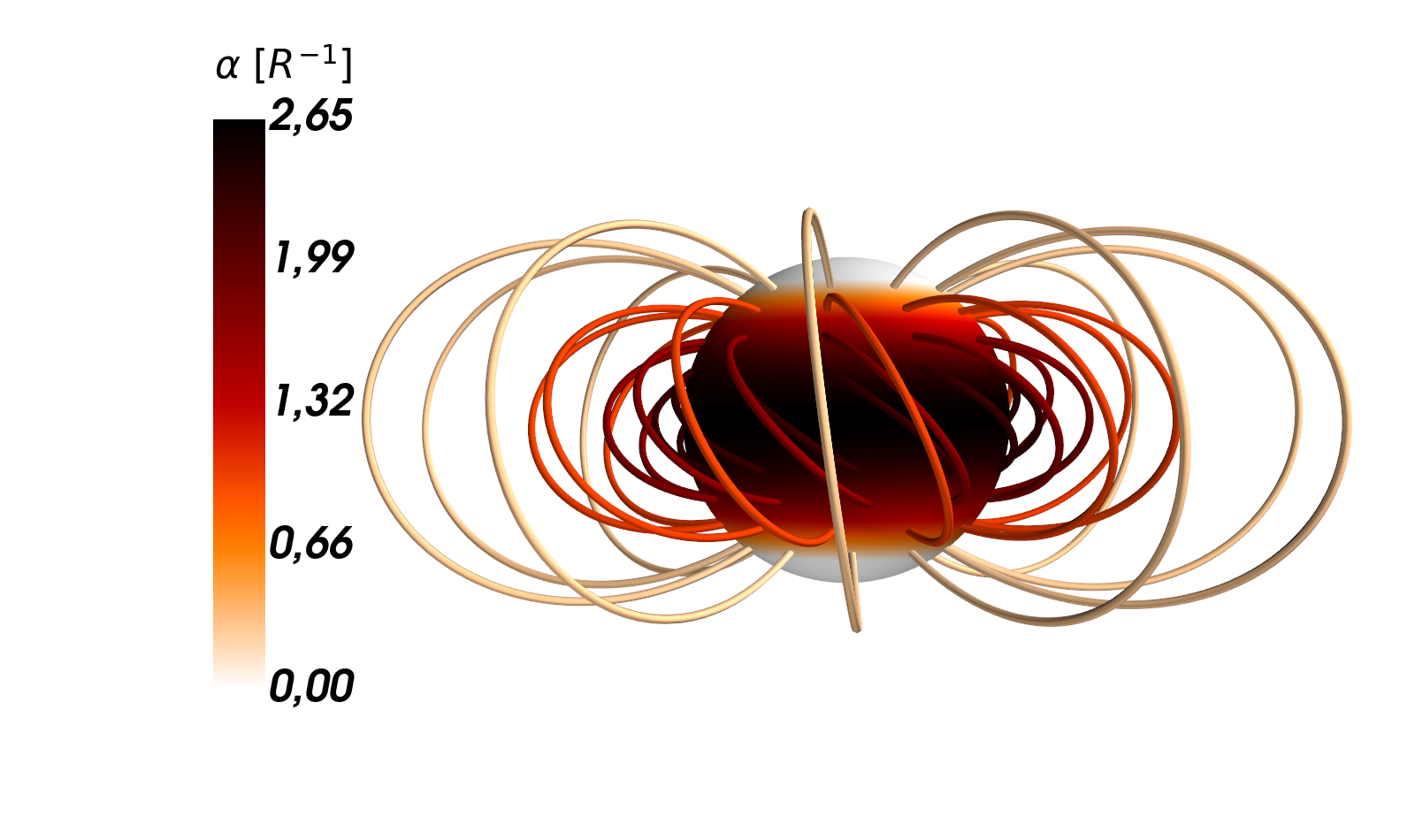}
    \caption{A reproduced example by the Grad-Rubin method, of an axisymmetric FF solution, akin to those studied in \citet{Akgun_Cerda_Miralles_Pons_2017}. Model parameters are: $P_c = 0.3405074, \sigma = 2$ and $s = 2$ and the solution shown corresponds to the lower energy branch.}
    \label{fig:ff2d_solution}
\end{figure}

As the two solutions have been calculated in different geometries and different grids, an interpolation of one of them in the grid of the other is needed in order to properly compare them. The FF2D solution has been calculated in a much finer, 2-D $400\times100$ grid, while the resolution of our grid is at most $160\times80\times160$ for this test. We choose, therefore, to interpolate the FF2D solution and evaluate it in our own, coarser, grid. The 3-D magnetic field configuration of the solution (as obtained by our code) is shown in Fig. \ref{fig:ff2d_solution}. 

Comparing the Grad-Rubin solution with the one obtained in \citet{Akgun_Cerda_Miralles_Pons_2017}, we see a converging pattern similar to the one shown in Fig. \ref{fig:convergence_dipole_tilted_energy} and \ref{fig:convergence_dipole_tilted_L2}. For the highest resolution used, the difference in energy is less than 1\% and in helicity less than 10\%. The $L_2$ norm of the difference of the $A_\phi$ component between the two solutions converges with a power-law index of $p=1.23$. All these metrics indicate a less effective convergence than in the previous test. This is understandable and can be attributed to two factors. First, the full Grad-Rubin method with currents (hyperbolic and elliptic part) has a complicated setup that involves iterations, integrations, interpolations and various other steps that can affect the convergence order. Second, in this test, we are comparing two numerical solutions produced by different numerical schemes, in different grids and an interpolation is performed between them.

\subsection{Test 3: Tilted axisymmetric FF Field}
In order to test our code in a configuration as general as possible, we make one more step by tilting the FF2D solution of section \ref{sec:test_ff2d}. This permits us to create a fully 3-D, current-filled magnetospheric configuration, with twisted lines crossing the coordinate axes. Although this is technically still an axisymmetric solution around some axis, it is not axisymmetric in our coordinate system and therefore can serve as a valid, general 3D test with currents. Energy and helicity are compared to the values of un-tilted FF2D solution, as they should not be affected by rotation. Again, we get a difference of less than 1\% for the energy and less than 10\% for the helicity. Due to technical reasons, it is easier to compare the $L_2$ norm of the difference in $B_\phi$ component for this case, instead of $A_\phi$. The convergence follows a power law of index $p=1.21$, simillar to the one for the un-tilted case.
    
\section{Results} \label{sec:results}

Observations of magnetars suggest that the radial current distribution over the NS's surface is localised into specific regions, the \textit{hotspots}, while the rest of the magnetosphere remains current-free. 
Current-filled lines emerge/close inside the hotspots, forming twisted magnetic loops. 
The resulting global magnetospheric configuration depends on the size, location, and shape of the spots, or the intensity of the current parameter $\alpha$, among other variables. 
In this section, we systematically explore the dependence of magnetospheric models on these parameters.

\subsection{Reference Model}\label{sec:reference_model}

We begin by discussing a characteristic example of the family of models considered in this work: a magnetic dipole, locally twisted due to the presence of two current inducing hotspots. 
Only field lines whose footprints lie inside the spots are twisted, while the rest of the magnetosphere is barely affected by the presence of the localised current loop and remains current-free and purely poloidal.
We impose a radial component of the magnetic field at the surface of the form
\begin{equation}
    b_S (\theta, \phi) = 2 \cos{\theta},
\end{equation}
corresponding to a dipole.
The boundary conditions are completed by giving values of $\alpha$ in the region where $B_q > 0$, by the relation
\begin{equation}
    \alpha_{S^+} (\theta, \phi) = \alpha_0 e^\frac{-(\theta - \theta_1)^2 - (\phi - \phi_1)^2}{2 \sigma ^2}.
\end{equation}
This corresponds to a Gaussian profile, centered at the point $(\theta_1, \phi_1)$ where $\sigma$ controls the width of the Gaussian. 
$\alpha_0$ determines the maximum of the parameter $\alpha$ (at the center of the Gaussian).  

\begin{figure}
    \centering
    \includegraphics[width=\columnwidth]{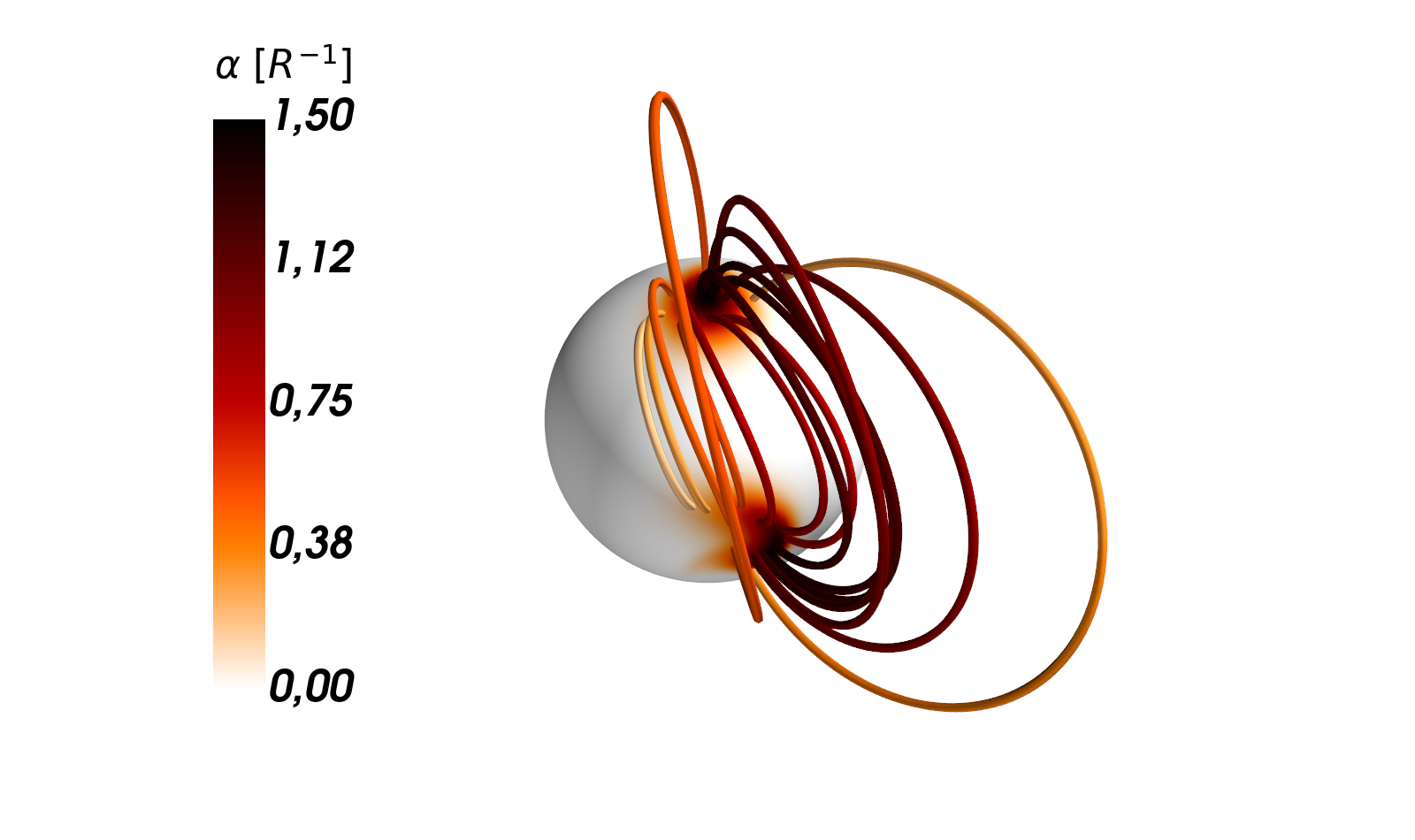}
    \caption{A selection of twisted field lines in a FF magnetosphere with two current-inducing hotspots. The colour code denotes the value of $\alpha$, which is constant along any particular field line. Model parameters are: $\alpha_0 = 1.5$, $\theta_1 = 45^\circ$, $\phi_1 = 180^\circ$ and $\sigma = 0.2$}
    \label{fig:model0_twisted_lines}
\end{figure}

\begin{figure}
    \centering
    \includegraphics[width=\columnwidth]{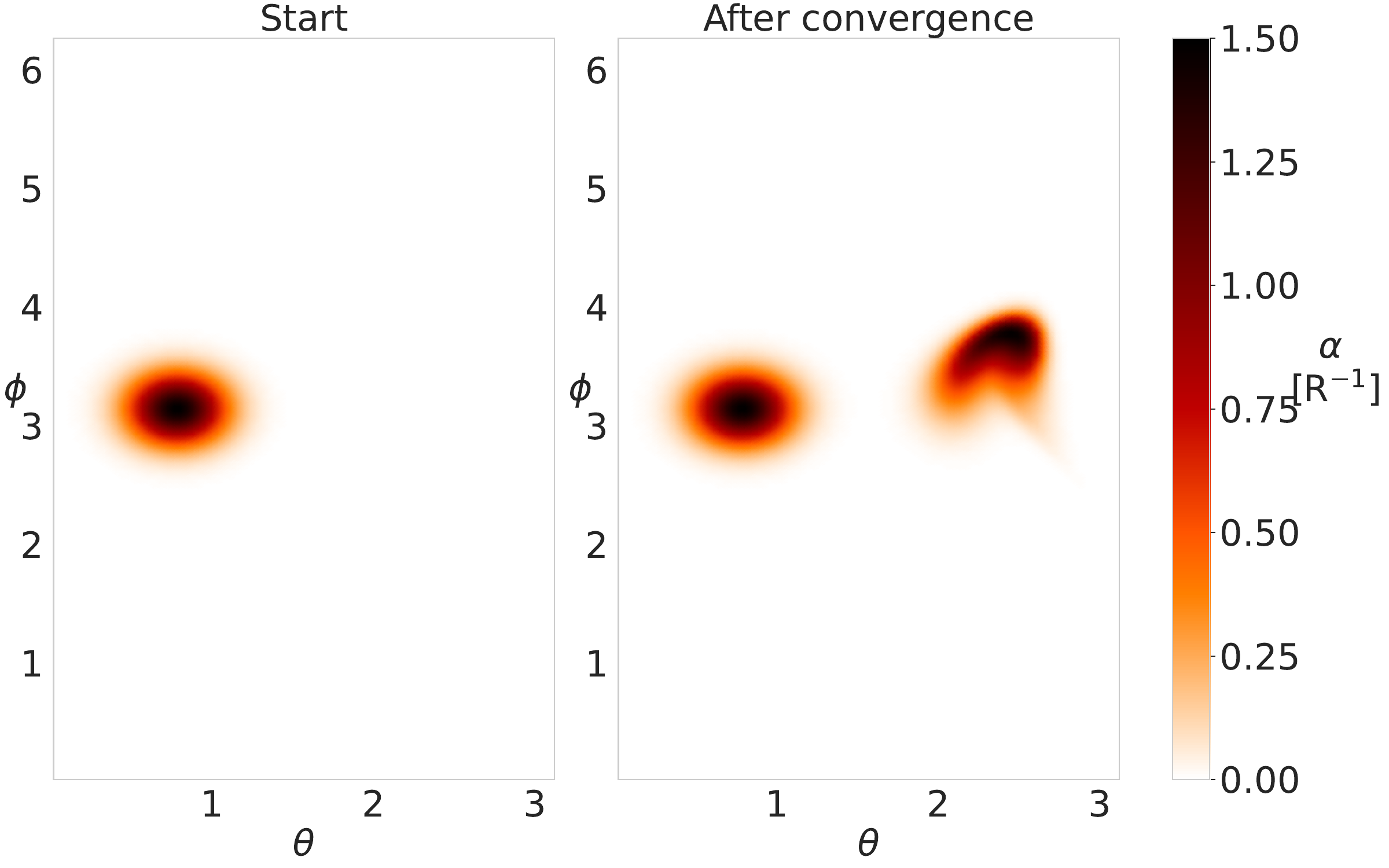}
    \caption{Surface values of $\alpha$ plotted in the $\theta-\phi$ plane. The left figure shows the prescribed values, only in the northern hemisphere where $B_q > 0$. The right figure shows the values after the solution has converged, corresponding to the real boundary. Model parameters are the same as in Fig. \ref{fig:model0_twisted_lines}.}
    \label{fig:model0_boundary}
\end{figure}

In Fig. \ref{fig:model0_twisted_lines} we show a selection of current-filled, twisted lines connecting the two hotspots.
The first immediately noticeable feature is the asymmetric "comma-like" shape of the hotspot residing in the southern hemisphere. 
We remind here that, to acquire a consistent solution, the surface values of $\alpha$ are given only for one polarity of the magnetic field. 
Thus, only one of the spots is fixed \textit{a priori} while the exact position and shape of the other is consistently determined.
Fig. \ref{fig:model0_boundary} compares the user-provided boundary condition for the first iteration (left) and the consistent boundary condition obtained after the solution has converged (right). We have tested that, using the comma-like spot as boundary condition we recover the Gaussian-like profile on the other side, showing that the numerical solution is consistent and this is not a numerical artefact.
Note that, interestingly, hotspots with a similar coma-like shape have been reported in the context of millisecond pulsars (\citet{Kalapotharakos_2021}), albeit created by a different mechanism (open field lines that cross the light cylinder, instead of closed twisted field lines).

\subsection{Twist}

The twist of a magnetic field line is defined mathematically as the integral of the torsion along the parametric curve $\bm{r}(\lambda)$ corresponding to the line:
\begin{equation}
    T_w = \int_\mathcal{C} \mathcal{T} d\lambda \label{eq:twist},
\end{equation}
where $\lambda$ is the length along the curve and the torsion $\mathcal{T}$ is given by the following formula, which makes use of the Frenet-Serret unit vectors: the tangent $\hat{\bm{\tau}}$, the normal $\hat{\bm{n}}$ and the binormal $\hat{\bm{b}} = \hat{\bm{\tau}} \times \hat{\bm{n}}$,
\begin{equation}\label{eq:torsion}
    \mathcal{T} = \frac{d\hat{\bm{n}}}{d\lambda} \cdot \hat{\bm {b}}.
\end{equation}
Calculating the Frenet-Serret unit vectors is straightforward in the case of magnetic field lines, since $\hat{\bm{\tau}} = \frac{\bm{B}}{B}$ and $\hat{\bm{n}} = \frac{\nicefrac{d\hat{\bm{\tau}}}{d\lambda}}{||\nicefrac{d\hat{\bm{\tau}}}{d\lambda}||}$. 
The problem with the above definition is that the consecutive differentiations introduce numerical errors and noise which can be fairly significant.
A simpler alternative definition \citep{BP2006} is directly related to the parameter $\alpha$
\begin{equation}
    T_\alpha = \frac{1}{2}\int_\mathcal{C} {\alpha} d\lambda \label{eq:twist_alt}
\end{equation}
and, since $\alpha$ is constant along a field line, the twist is simply given by
\begin{equation}
    T_\alpha = \alpha L.
\end{equation}
We refer the interested reader to Appendix C in \cite{LK2016} for an excellent and detailed discussion of the differences between the two definitions. 

\begin{figure}
    \centering
    \includegraphics[width=\columnwidth]{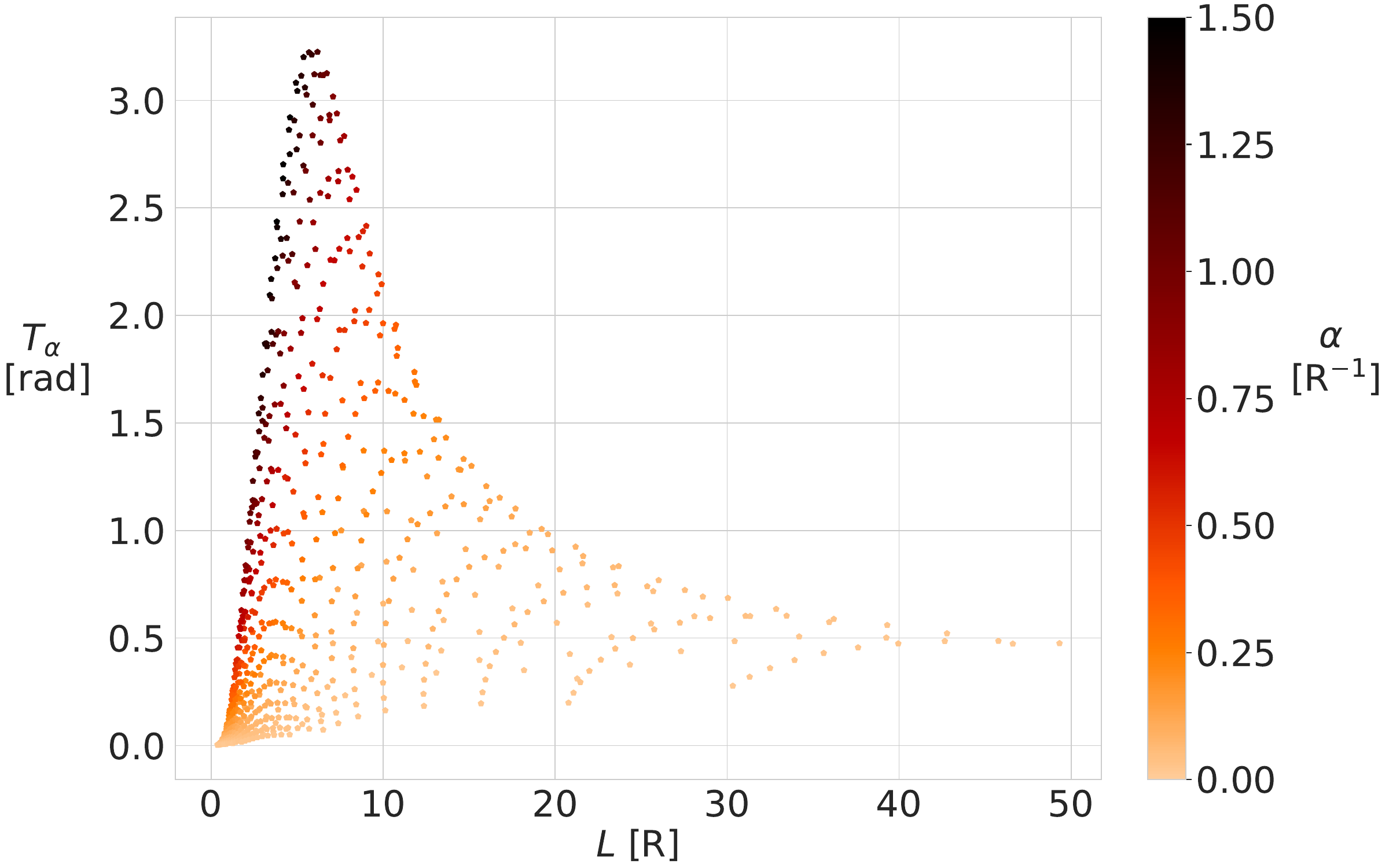}
    \caption{Twist of the magnetic field lines as a function of the total length. Each point corresponds to one field line of the reference model discussed in section \ref{sec:reference_model}. The colour denotes the value of $\alpha$ labelling the line. Twist is measured in radians, using the definition \eqref{eq:twist_alt}. Lines with high $\alpha$ tend to have larger twists.}
    \label{fig:model0_twist_alt}
\end{figure}

Fig. \ref{fig:model0_twist_alt} 
shows the distribution of twists of a sample of field lines with footprints on the hotspot as calculated as a function of the total length $L$ of the line in units of the stellar radius $R$ (Eq. \eqref{eq:twist_alt}). The colour scale indicates the value of $\alpha$ that labels the line. 
Lines with footprints close to the centre of the spot are shown with a deep red colour and have higher values of the twist (forming the peak in the plot). A faded orange colour corresponds to lines emerging from the periphery of the hotspot, where $\alpha$ is relatively low and the line is only slightly twisted, but these are generally longer lines (forming the tail of the plot). The Very long and twisted lines should be unstable beyond certain limit, as illustrated by the Figure, where the absence of points occupying the upper right corner reflects the fact that a line cannot have simultaneously large values of $\alpha$ and $L$.


\subsection {Dependence on $\alpha_0$}

\begin{figure*}
    \centering
    \begin{subfigure}[b]{0.33\textwidth}
        \centering
        \includegraphics[width=\textwidth]{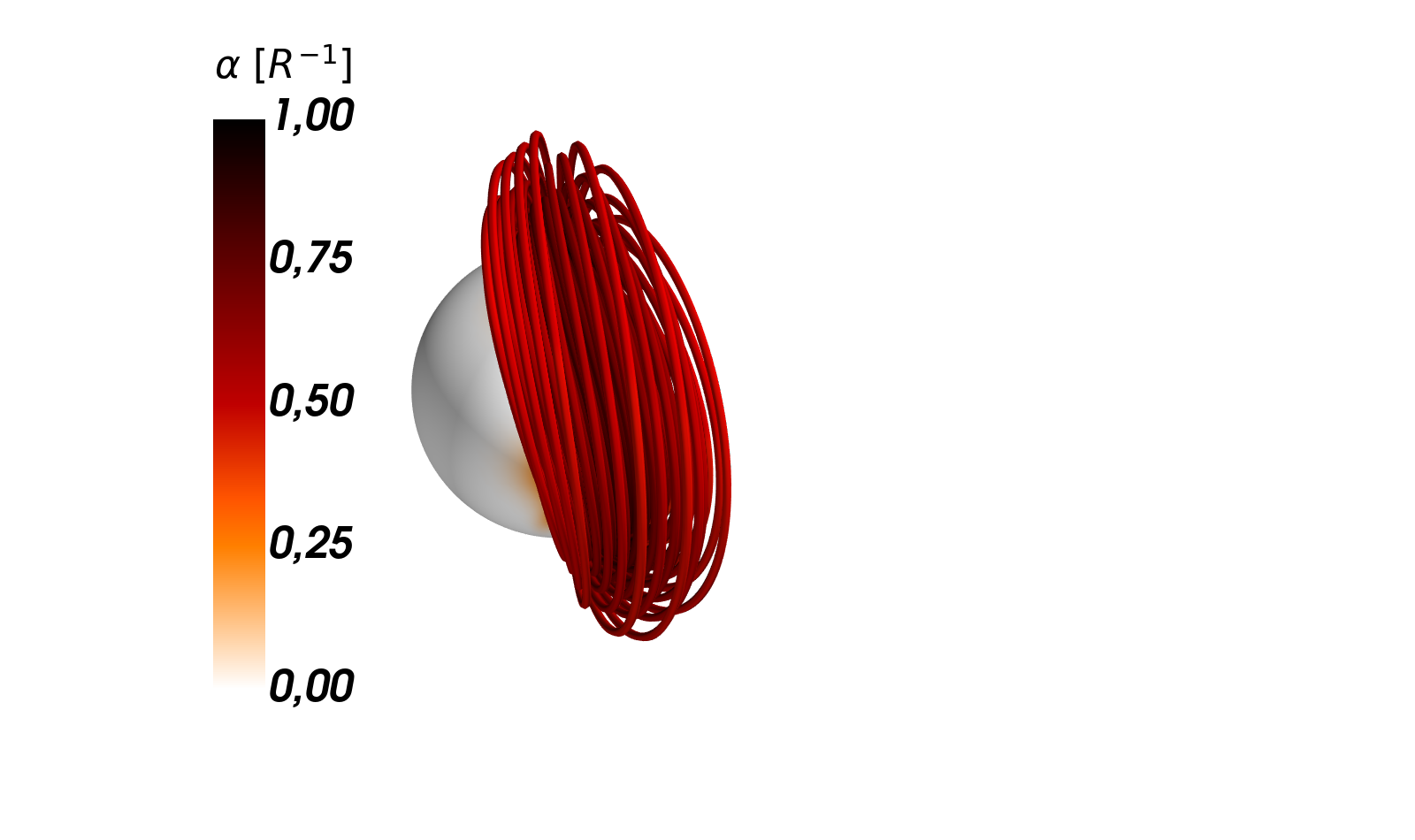}
        \caption{$\alpha_0 = 1.0$}
        \label{fig:sweep_a0_samples_1}
    \end{subfigure}
    \hfill
    \begin{subfigure}[b]{0.33\textwidth}
        \centering
        \includegraphics[width=\textwidth]{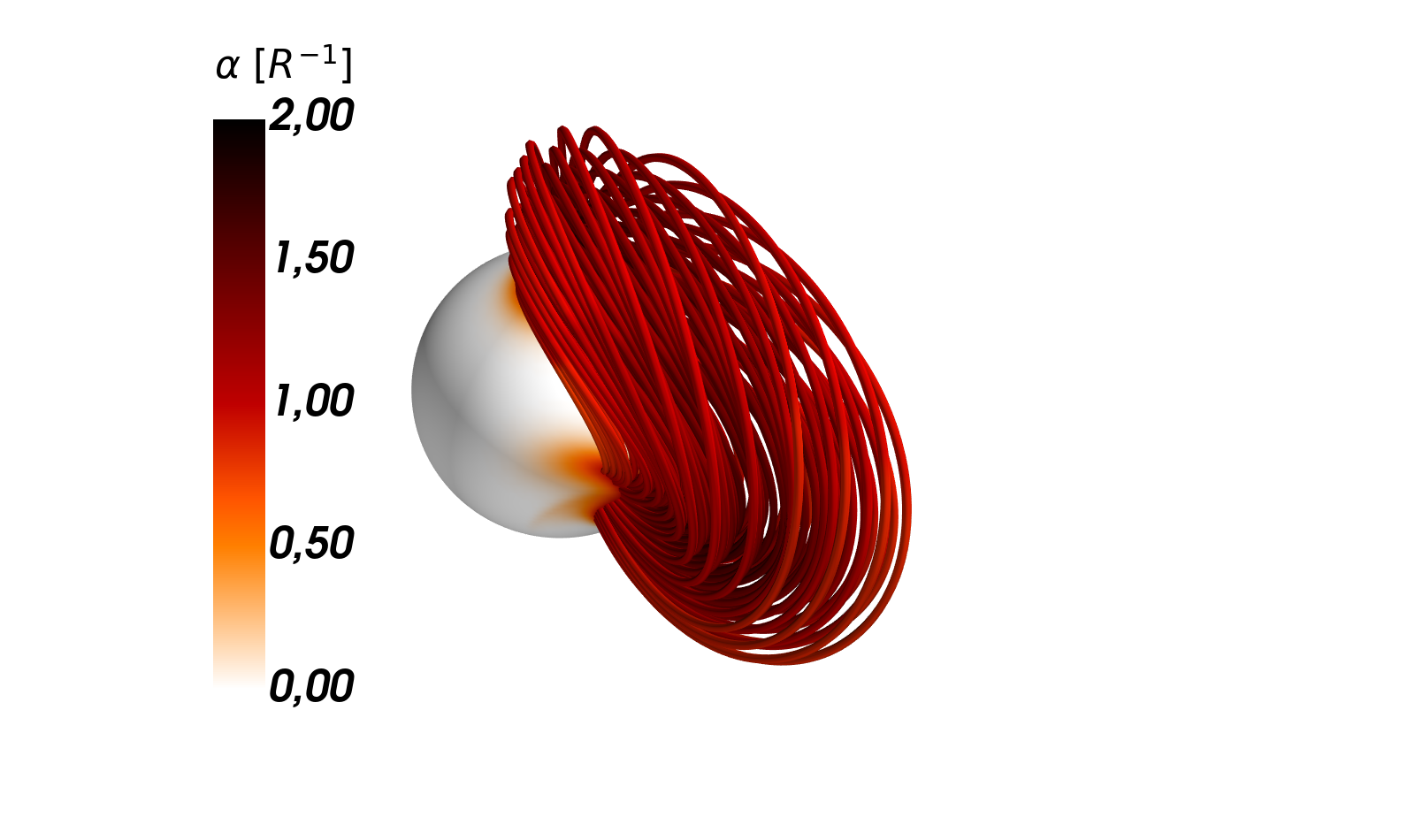}
        \caption{$\alpha_0 = 2.0$}
        \label{fig:sweep_a0_samples_2}
    \end{subfigure}
    \hfill
    \begin{subfigure}[b]{0.33\textwidth}
        \centering
        \includegraphics[width=\textwidth]{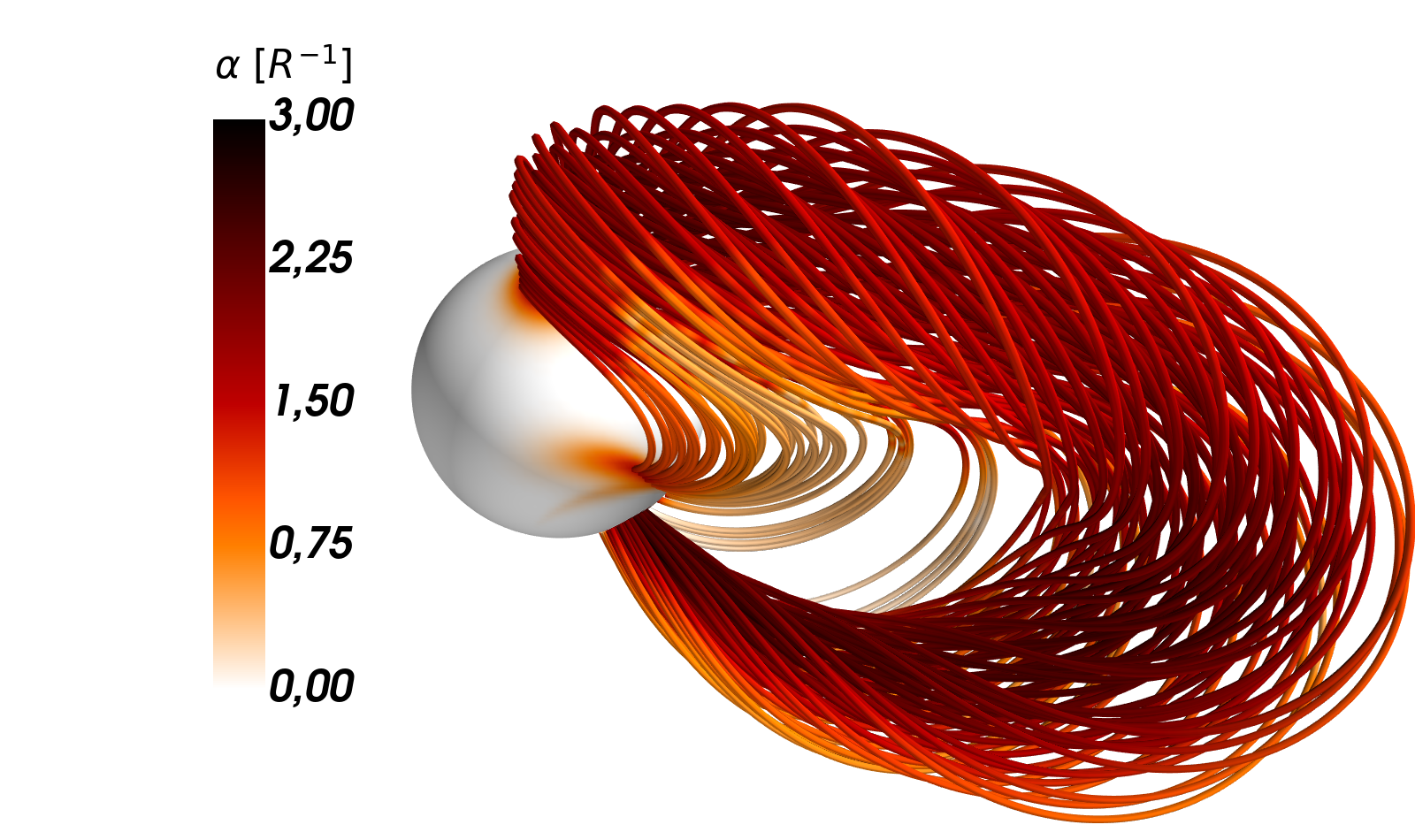}
        \caption{$\alpha_0 = 3.0$}
        \label{fig:sweep_a0_samples_3}
    \end{subfigure}

    \begin{subfigure}[b]{0.33\textwidth}
        \centering
        \includegraphics[width=\textwidth]{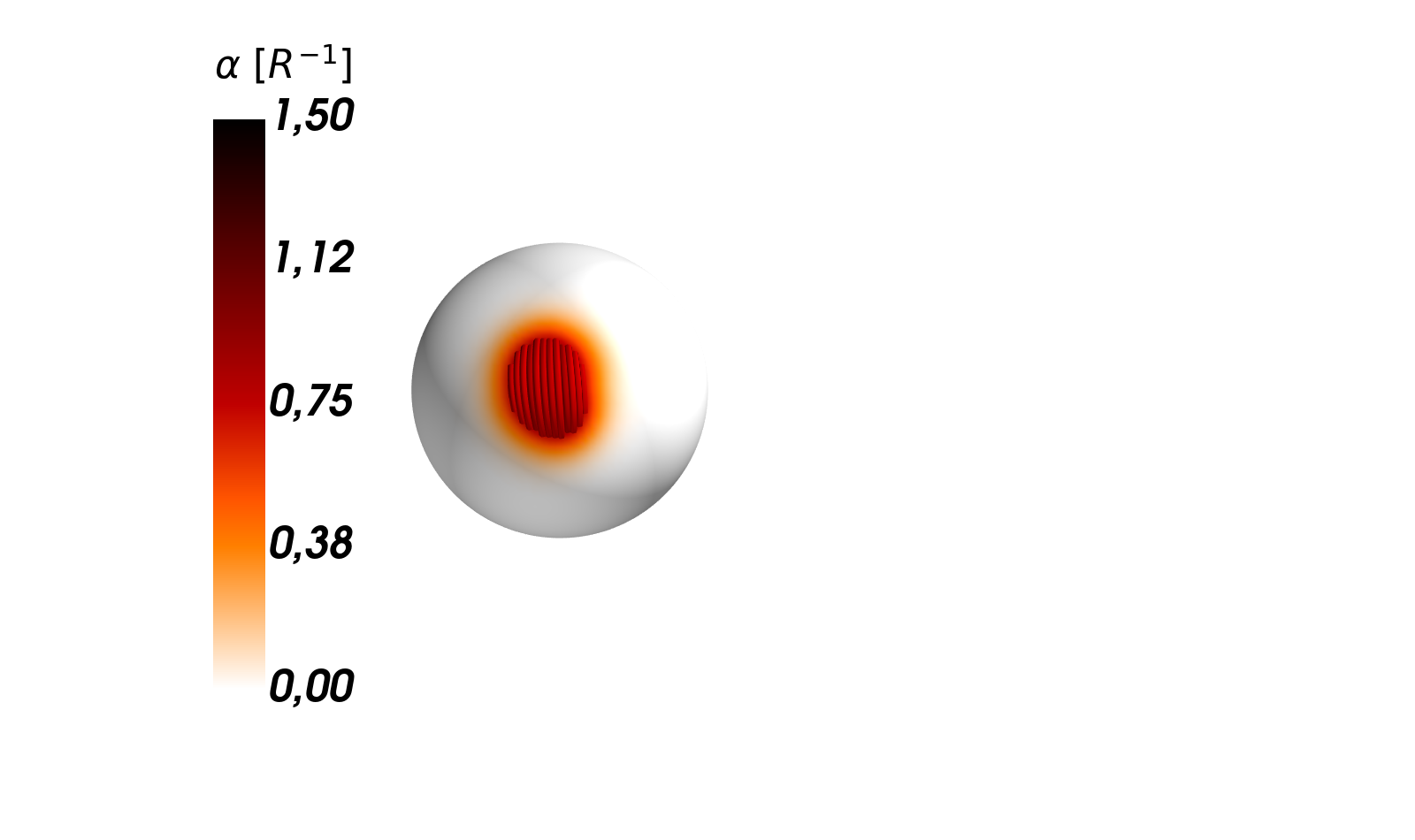}
        \caption{$\theta_1 = 85^\circ$}
        \label{fig:sweep_theta1_samples_1}
    \end{subfigure}
    \hfill
    \begin{subfigure}[b]{0.33\textwidth}
        \centering
        \includegraphics[width=\textwidth]{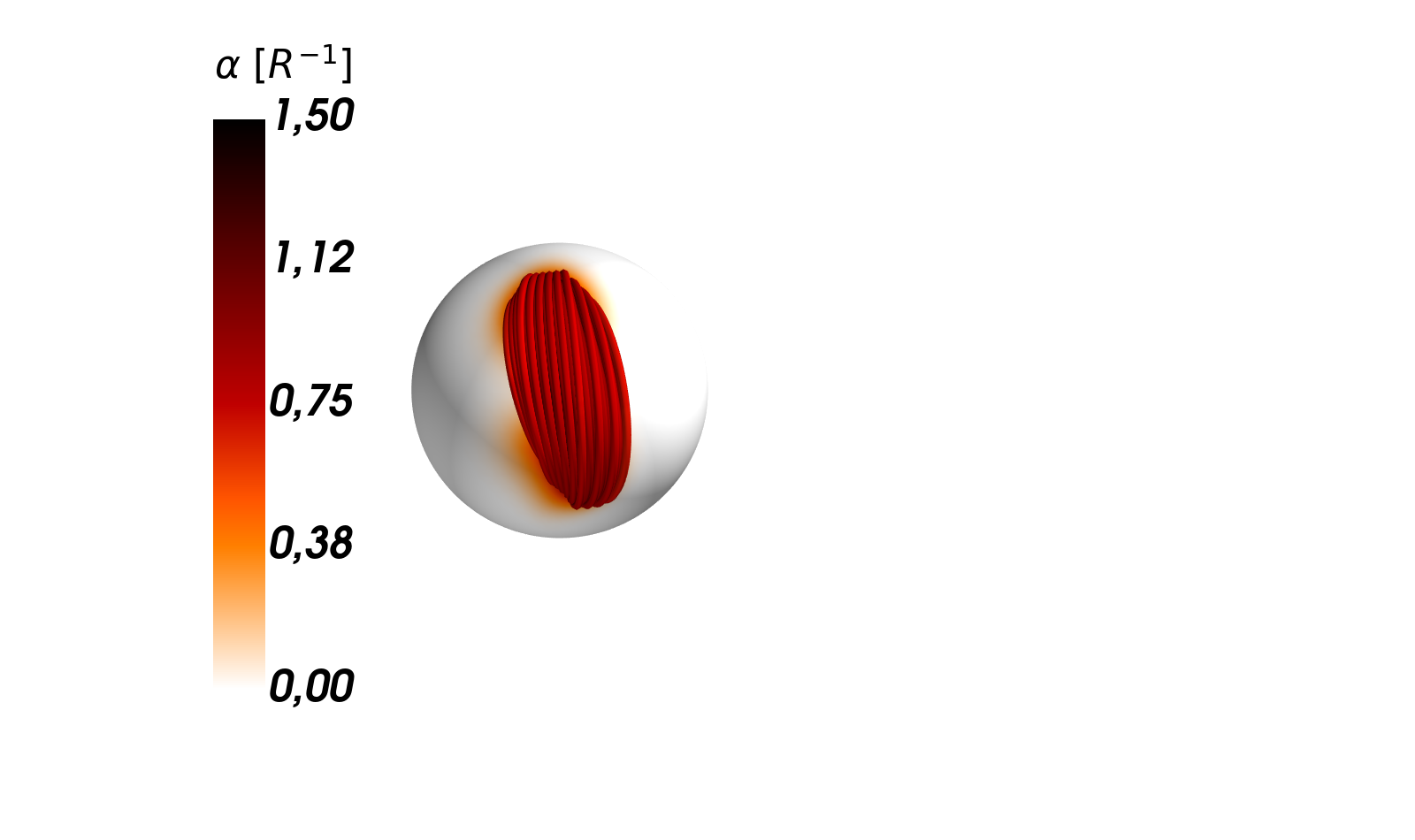}
        \caption{$\theta_1 = 60^\circ$}
        \label{fig:sweep_theta1_samples_2}
    \end{subfigure}
    \hfill
    \begin{subfigure}[b]{0.33\textwidth}
        \centering
        \includegraphics[width=\textwidth]{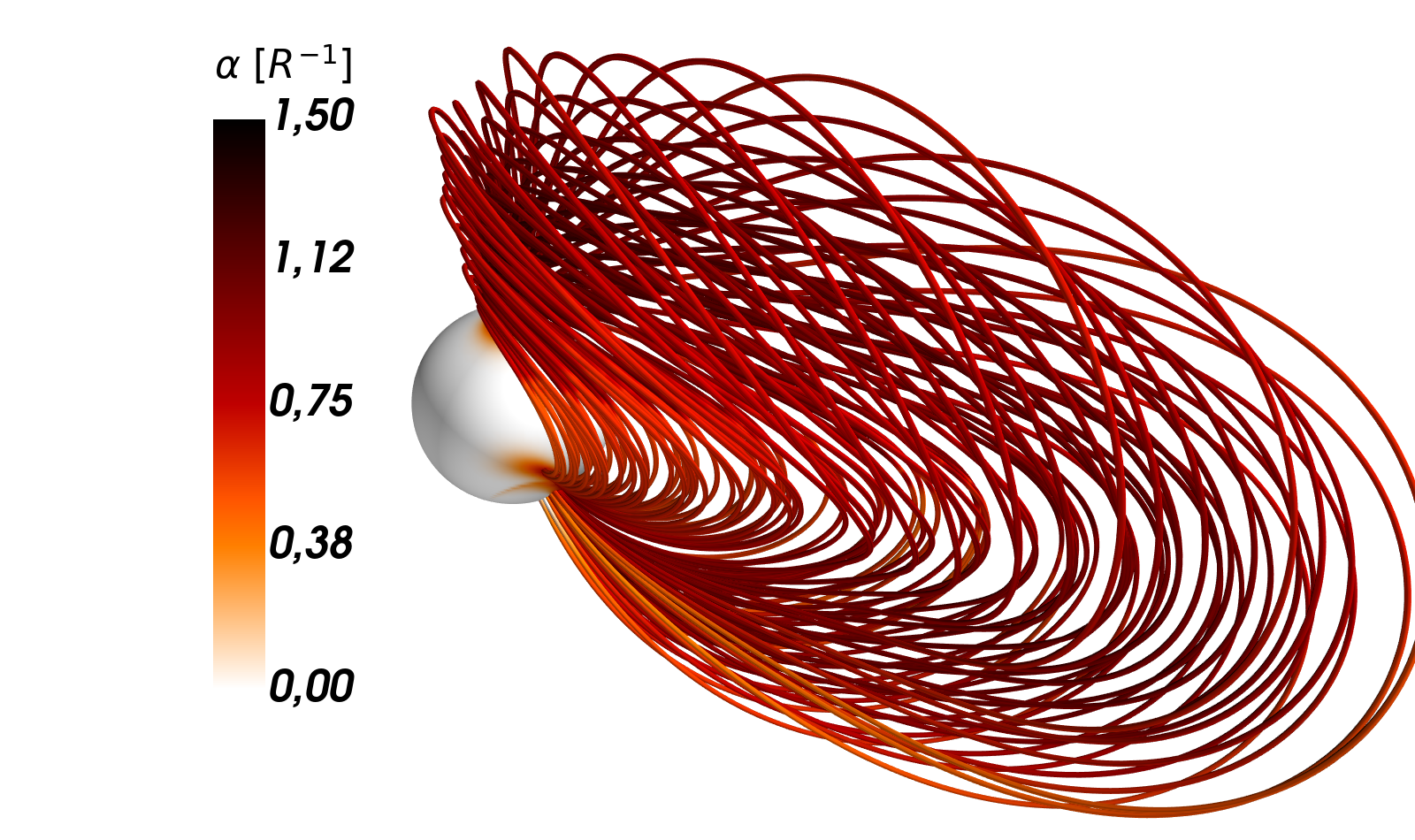}
        \caption{$\theta_1 = 35^\circ$}
        \label{fig:sweep_theta1_samples_3}
    \end{subfigure}
    \hfill
    
    \begin{subfigure}[b]{0.33\textwidth}
        \centering
        \includegraphics[width=\textwidth]{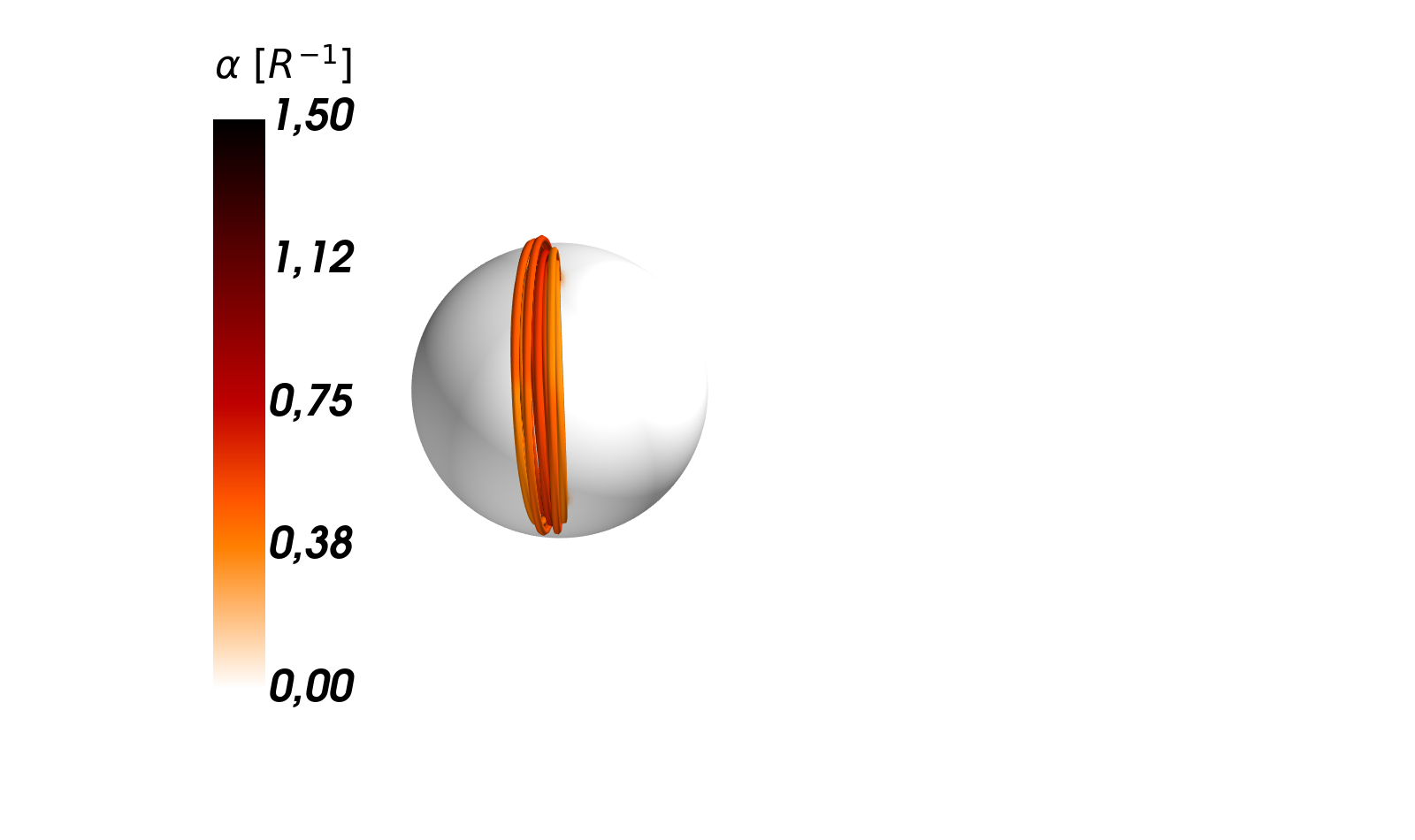}
        \caption{$\sigma = 0.05$}
        \label{fig:sweep_sigma_samples_1}
    \end{subfigure}
    \hfill
    \begin{subfigure}[b]{0.33\textwidth}
        \centering
        \includegraphics[width=\textwidth]{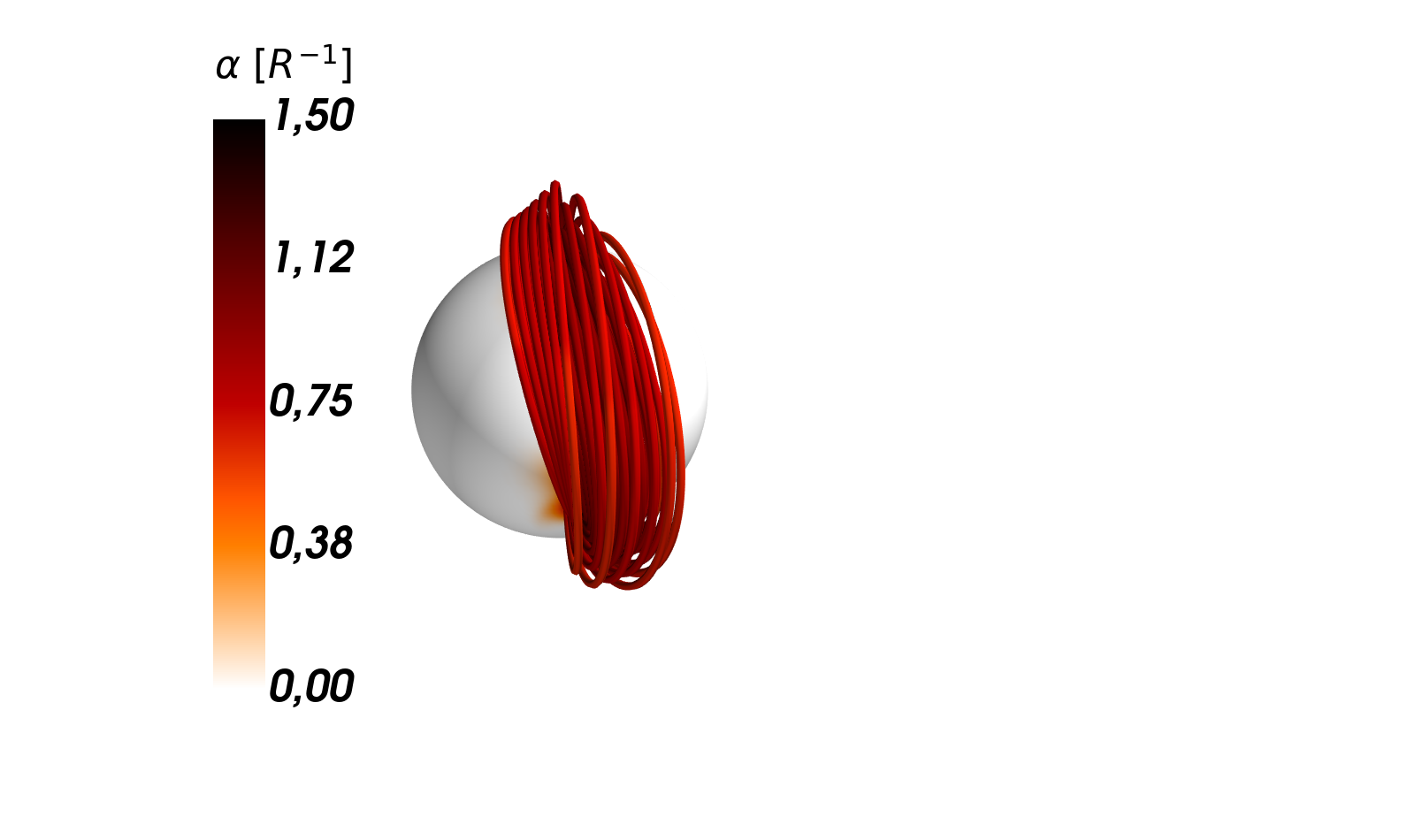}
        \caption{$\sigma = 0.15$}
        \label{fig:sweep_sigma_samples_2}
    \end{subfigure}
    \hfill
    \begin{subfigure}[b]{0.33\textwidth}
        \centering
        \includegraphics[width=\textwidth]{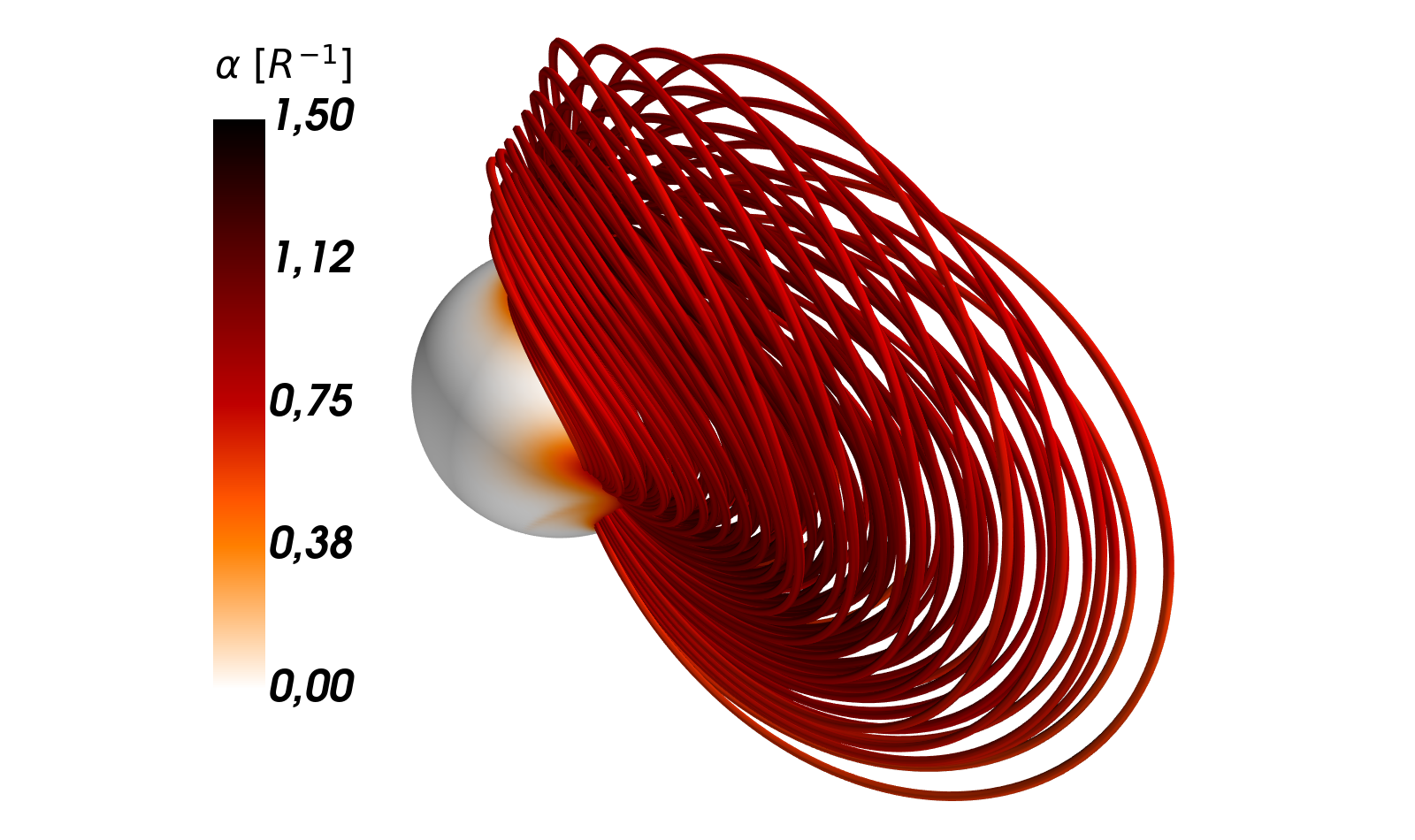}
        \caption{$\sigma = 0.25$}
        \label{fig:sweep_sigma_samples_3}
    \end{subfigure}

    \caption{Samples of magnetospheres with different values of $\alpha_0$, $\theta_1$ and $\sigma$. In each row, only one parameter varies while the others remain fixed to the reference model values. In each column, from left to right, the magnetosphere is farther away from the potential current-free solution. The magnetic field is more twisted for strong currents (high $\alpha_0$), large hotspot separation (small $\theta_1)$ and large hotspot area (high $\sigma$). For clarity, only lines with $\alpha > 0.5 \alpha_0$ are shown.}
    \label{fig:sweep_parameters_samples}
\end{figure*}

The parameter $\alpha_0$ controls the intensity of the current. 
As its value increases, the twist of the lines threaded by currents becomes stronger. In the first row of 
Fig. \ref{fig:sweep_parameters_samples} we show samples of solutions with different values of $\alpha_0$. In order to quantify the effect of $\alpha_0$ on the magnetosphere, we examine the dependence of various important physical quantities on it.

We define the normalised excess energy as
\begin{equation}\label{eq:excess_energy}
    E_e = \frac{E-E_d}{E_d},
\end{equation}
where $E_d$ is given by Eq. \eqref{eq:dipole_energy}. Considering that the dipole is the lowest energy configuration, this quantity gives an estimate of the available energy budget if the magnetosphere is eventually completely untwisted and relaxed.

\begin{figure*}
    \centering
    \includegraphics[width=\textwidth]{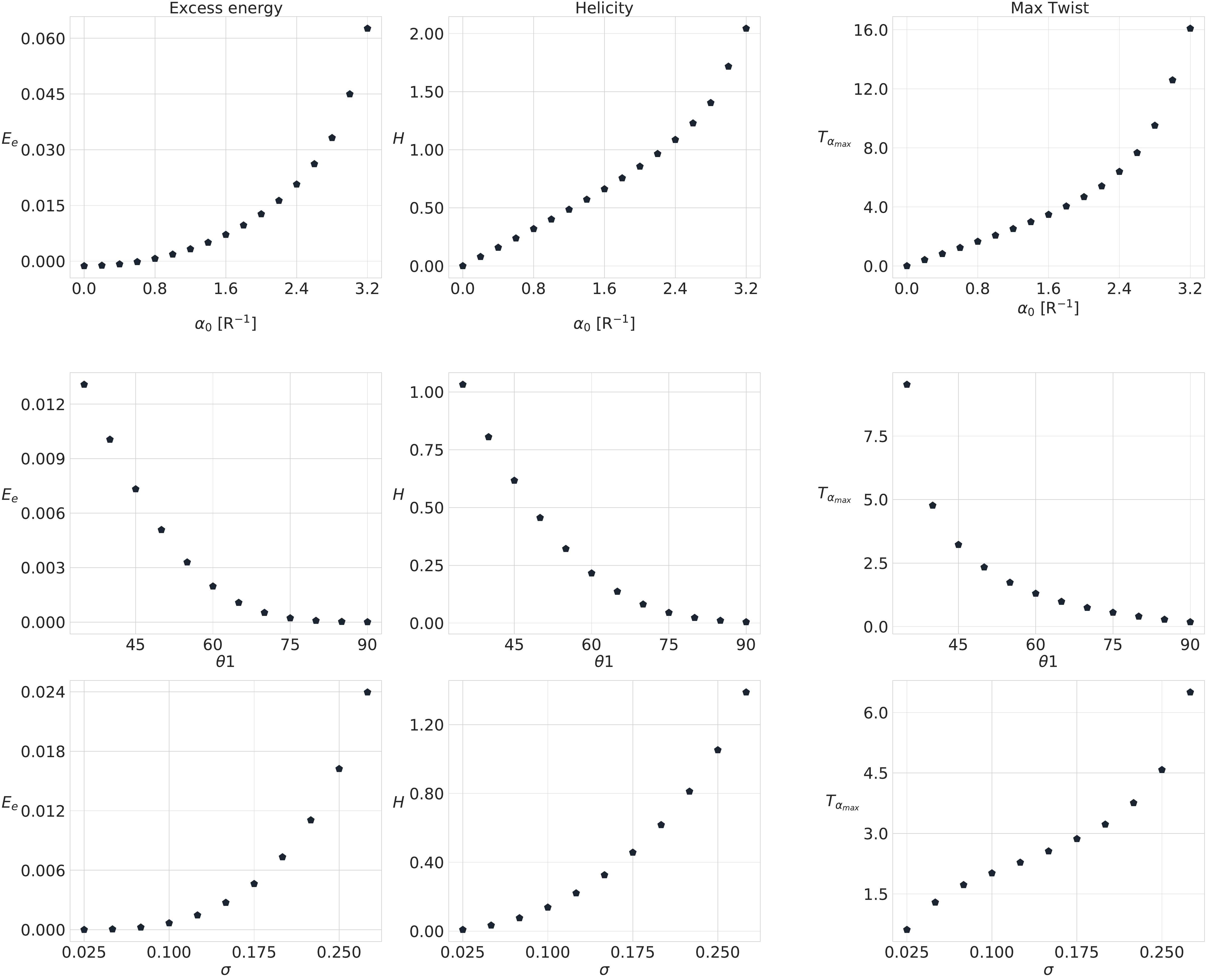}
    \caption{Dependence of the normalised excess energy $E_e$, the helicity $H$ and the maximum twist $T_{\alpha_{max}}$ on the parameters $\alpha_0$ (first line), $\theta_1$ (second line) and $\sigma$ (third line).}
    \label{fig:dependence_on_parameters}
\end{figure*}

The first row of Fig. \ref{fig:dependence_on_parameters} shows the dependence of the normalised excess energy $E_e$, the helicity $H$ and the maximum twist $T_{\alpha_{max}}$ (as calculated by Eq. \eqref{eq:twist_alt}) on $\alpha_0$. 
As expected, all quantities increase as we get further and further away from the potential solution ($\alpha_0 = 0$). 
Beyond a value of $\alpha_0 \sim 3$, the numerical code is unable to converge to a solution. 
 The non existence of a solution can be a flag for the triggering of instabilities in the magnetosphere, as it becomes unable to support such intense currents and strong twists. 
 Looking at Fig. \ref{fig:sweep_a0_samples_3}, one can clearly see that magnetospheres with such high values of $\alpha$ threading sufficiently long lines are highly distorted and may not be stable. 

\subsection{Dependence on $\theta_{1}$}

The parameter $\theta_1$ determines the position of the hotspot. 
A small value means that the spot is located close to the pole. 
If the initial, untwisted magnetic field is a dipole, it essentially controls the distance between hotspots and the total volume of the coronal flux tube (the farther apart the footprints, the longer the fluxtube length and the larger the volume).
Note that it is meaningless to consider values of $\theta_1 > 90 ^ \circ$ for our particular untwisted magnetic field. 
Samples of solutions with different values of $\theta_1$ are shown in the second row of Fig. \ref{fig:sweep_parameters_samples}. 
The second row in Fig. \ref{fig:dependence_on_parameters} shows the dependence of $E_e$, $H$ and $T_{\alpha_{max}}$ on $\theta_1$. 
All three quantities increase as the hotspot's location approaches the magnetic pole. 
This is the expected behaviour, because the size of the current-filled region becomes larger. 
Changing the value of $\theta_1$ seems to have a lesser impact on energy and helicity than increasing the intensity of the current $\alpha_0$. 
However, it has a significant impact on the maximum twist. 
When current-filled lines, even with small values of $\alpha$ approach the pole, their twist is high due to their length and is difficult to acquire solutions.
This can be explained by the fact that our assumptions for the FF regime demand that the region containing currents cannot extend up to large distances.

\subsection{Dependence on $\sigma$}

The parameter $\sigma$ determines the area of the hotspot. 
We only consider cases with common $\sigma$ for the $\theta$ and $\phi$ directions, which means that the hotspot is symmetric.
A bigger value of $\sigma$ means that more field lines are threaded with current. This affects the total volume of the current-filled region but also its extent, since a big $\sigma$ means that lines close to the poles can have current.
In the third row of \ref{fig:sweep_parameters_samples} we show samples of solutions with different values of $\sigma$ while in the third row of Fig. \ref{fig:dependence_on_parameters} we show how energy, helicity and maximum twist alter with increasing the size of the hotspot. 
As expected, models with large values of $\sigma$ can store more energy, since the total volume of the twisted region increases.

\section{Discussion} \label{sec:discussion}

In the models presented in the previous section, the excess energy stored in the magnetosphere can reach values up to $E_e \sim 6\%$ of the energy of the potential solution. 
This is the maximum available energy that a twisted magnetosphere can release in a catastrophic event.
The magnetic energy of a purely dipolar field is 
\begin{equation}
    E_d = 3.3 \times 10^{45} \text{ erg} \ B_{14}^2 \left(\frac{R}{10 \text{ km}}\right)^3,
\end{equation}
where $B_{14}$ is the field in units of $10 ^ {14}$ G.
If one considers typical values for a magnetar, i.e. a star of radius 10 km and a magnetic field of the order of $10^{13} - 10^{15}$ G, $E_e$ lies in the range of $2 \times 10^{42} \text{ erg} < E_e < 2 \times 10^{46} \text{ erg}$, well in line with the typical
energetics of magnetar flares and outbursts, but perhaps falls a little short in comparison to the highly energetic giant flares.

Note that an exhaustive analysis of the entire parameter space lies outside the scope of this work. 
It is possible that a certain combination of $\alpha_0, \theta_1$ and $\sigma$, or even a different current profile would produce higher values of $E_e$.
In particular, if we associate giant flares to very large scale global events, a configuration with smaller $\theta_1$ and large $\alpha$ would be indeed consistent with the most extreme cases.

However, it is highly unlikely that the results could differ drastically from what has been presented here, given the fact that finding more extreme solutions becomes increasingly more  difficult. So we do not expect that hypothetical hypergiant or supergiant flares be possible. Observed giant flares seem to be already at the edge of the possibilities consistent with these family of models.

Concerning the twist, it tends to reach values significantly higher than the maximum values reported for axisymmetric magnetospheres (about $~1.5$ rads). 
This can be partly attributed to the fact that, in 3D,
lines can freely twist and bend and even change direction without the restrictions imposed by axisymmetry. 

An additional question relevant to the interpretation of magnetar observations is how these models differ from pure current-free magnetospheres in terms of the emission spectrum. A first relevant issue is that
currents flowing through the star surface at the footprints of the twisted magnetic lines can heat the surface (hence the name hotspots). 

\begin{figure}
     \centering
     \includegraphics[width=\columnwidth]{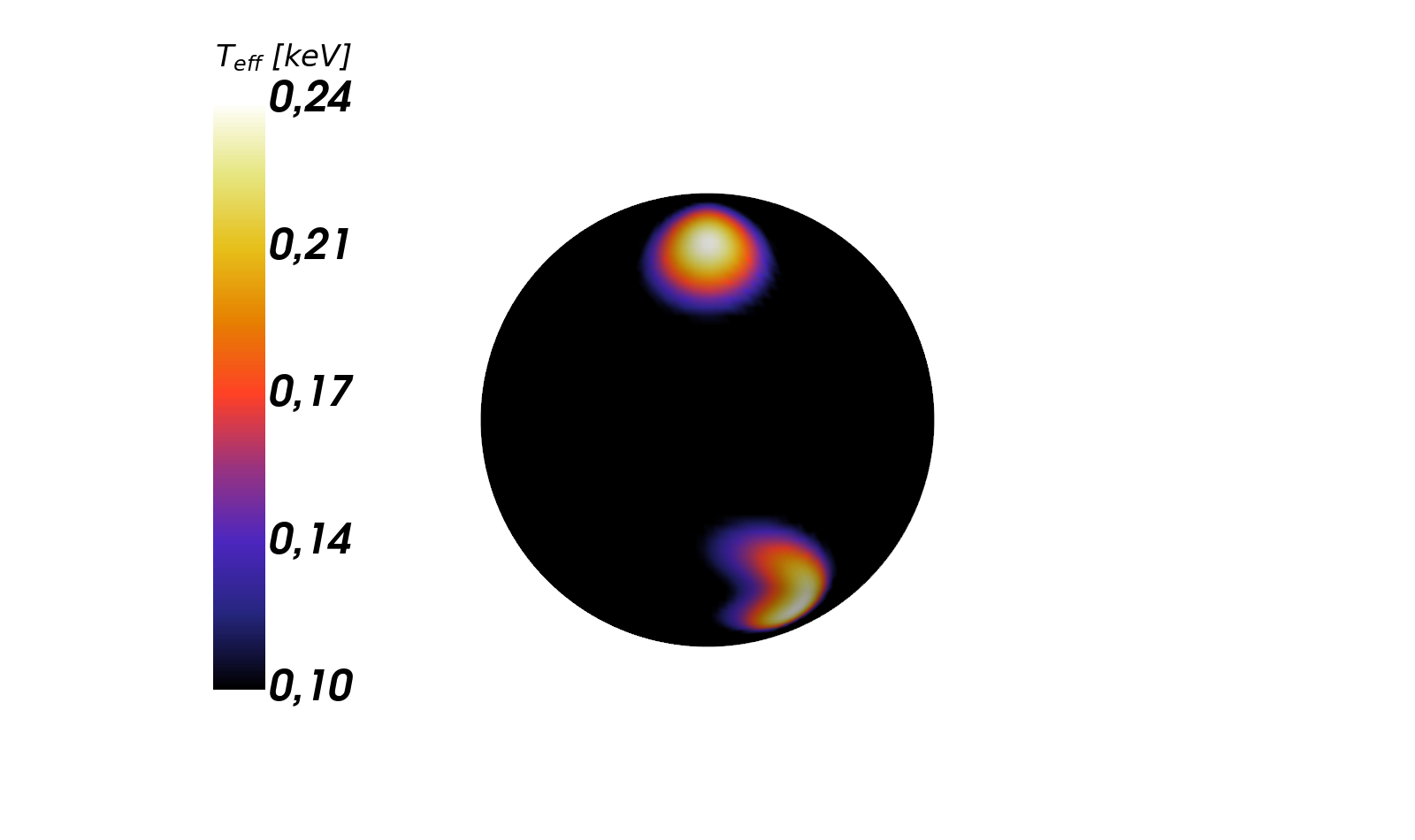}
     \caption{Effective temperature of the hotspots at the surface of the star as calculated by Eq. \eqref{eq:T_eff} for the reference model of section \ref{sec:reference_model}. The colorbar shows the values of $T_{eff}$ in keV units. The lowest value of the colorbar corresponds to the typical surface temperature of a neutron star in quiescence.}
     \label{fig:T_eff}
 \end{figure}

One can estimate the temperature of the hotspot following the arguments in \cite{Akgun_Cerda_Miralles_Pons_2018b}.
Assuming that the current is predominantly dissipated in a thin layer below the star surface (the region with highest resistivity) and the energy released is radiated away as black body radiation from the surface of the star, the effective temperature can be calculated from \citep[see Eq. (11) from][]{Akgun_Cerda_Miralles_Pons_2018b}
\begin{equation}\label{eq:T_eff}
    T_{eff} \simeq 0.2 \text{ keV}  \left( \frac{\Delta r}{ 1 \text{ m}} \right)^{\frac{1}{4}} \left( \frac{\eta}{10^4 \text{ cm}^2/\text{s}} \right)^{\frac{1}{4}} 
    B_{14}^{\frac{1}{2}} 
    \left( \alpha R \right)^{\frac{1}{2}}
\end{equation}
Considering $\Delta r$ and $\eta$ are of order 1 in the units given in the above expression, a magnetic field of $B_{14} \approx 1-10$ G and values of $\alpha R$ in the range $1 - 3$, the effective temperature of the hotspot lies in the range of $0.1 \text{ keV}< T_{eff} < 0.6 \text{ keV}$. 
Although this is a rough estimate, these values are on par with observations \citep{Zelati_2017, Younes_et_al_2022}. 
Furthermore, our Gaussian profiles of $\alpha$ at the hotspot are consistent with the double BB models and the umbral/penumbral shape for the pulsed X-ray emission reported in \citet{Younes_et_al_2022}, with the hotter component coming from the centre of the Gaussian (say within 1 $\sigma$) and the colder one from the periphery, where $\alpha$ and hence $T_{eff}$ have decayed. 
Fig. \ref{fig:T_eff} shows a surface map of $T_{eff}$ that corresponds to the reference model of section \ref{sec:reference_model}, in real units.
Currents in the magnetosphere are not permanent, but rather gradually dissipate. The diffusion timescale can be approximated by 
\begin{equation}
    \tau \simeq 3 ~yr \left( \alpha R \right)^{-2} \left( \frac{\eta}{10^4 \text{ cm}^2/\text{s}} \right)^{-1}.
\end{equation}
That means that stronger twists ($\alpha R > 1$) tend to live for a few months up to several years, while weaker twists last longer. This is in line with expectations from outburst observations.

 Ideally, a joint simulation of the interior and the exterior of the star, where the magneto-thermal evolution of the field in the crust is coupled with a FF magnetosphere, would give us the exact location and shape of the hotspots, as well as a consistent way of estimating its temperature. Such extension is out of our present scope and reserved for future work. 
 
\section{Conclusion} \label{sec:conlcusion}
 In this work, we have presented a family of 3D FF twisted models for magnetar magnetospheres.
 We have considered a dipolar magnetic field and a localised current with a Gaussian profile at the surface of star. 
 Our goal is to emulate the behaviour of the magnetosphere in the presence of current-inducing hotspots.
 We have studied the effect of the parameters controlling the strength and geometry of the current-filled region on physical quantities such as the energy, helicity and twist of the magnetic field, the effective surface temperature of the hotspot and the diffusion timescale.
 Our results show good agreement with the observations and the expected properties of real magnetars.
 The next step is to consider a coupling of the magnetospheric solution with the internal magneto-thermal evolution so that the hotspot properties (values of $\alpha$ and geometry) could be determined self-consistently.
\section*{Acknowledgements}

We acknowledge support from the Generalitat Valenciana (PROMETEO/2019/071) and the AEI grants PGC2018-095984-B-I00, PID2021-125485NB-C21 and PID2021-127495NB-I00. JAP acknowledges the Alexander von Humboldt Stiftung for a Humboldt Research Award. PCD acknowledges support from the {\it Ramon y Cajal} programme of the AEI (RYC-2015-19074).

\section*{Data Availability}

All data produced in this work will be shared on reasonable request to the corresponding author.



\bibliographystyle{mnras}
\bibliography{Bibliography} 




\appendix

\section{Poloidal-Toroidal Decomposition} \label{app:poloidal_toroidal_decomposition}

A commonly used practice to deal with the magnetic field in curvilinear coordinates is to decompose it in a poloidal and a toroidal component and express these components in terms of two scalar functions. Through the literature, various approaches can be found, regarding what the scalar functions represent, as well as notation for the quantities introduced. In this section, we go through the poloidal-toroidal decomposition equations in our notation (see \citet{Geppert_Wiebicke_1991} for a more detailed treatment of the subject).

The principle behind this process is to introduce two scalar functions $\Phi (\bm{r}), \Psi (\bm{r})$ and split the magnetic field to a purely poloidal and a purely toroidal part as follows
\begin{align}
    \bm{B}_{p} &= \nabla \times (-\bm{r} \times \nabla \Phi) \label{eq:B_pol}, \\ 
    \bm{B}_{t} &= -\bm{r} \times \nabla \Psi  \label{eq:B_tor}, 
\end{align}
where $\bm{r} = r\hat{\bm{r}}$ is the position vector and we are working in spherical coordinates $(r,\theta, \varphi)$.
From the definition of $B_{p}$ one can readily deduce that the toroidal component of the vector potential is given by 
\begin{equation}\label{eq:toroidal_vector_potential}
    \bm{A}_t = -\bm{r} \times \nabla \Phi.
\end{equation}
Any solenoidal field can be unambiguously represented from equations \eqref{eq:B_pol}, \eqref{eq:B_tor}. Expanding Eq. \eqref{eq:B_pol} with the help of vector calculus identities gives us
\begin{equation}
    \bm{B}_{p} = -\bm{r} \nabla^2 \Phi + \nabla \frac{\partial}{\partial r} (r\Phi)\label{eq:B_pol_2}.
\end{equation}
Toroidal fields are purely angular, they do not have a radial component. All the information about the radial magnetic field is incorporated in $B_{p}$. If we isolate only the radial component of Eq. \eqref{eq:B_pol_2} we end up with an equation that completely describes the radial magnetic field
\begin{equation}
    B_r = -r \nabla^2 \Phi + r \frac{\partial^2}{\partial r^2} \Phi.
\end{equation}
One can see that the second term in the right-hand side in the above equation cancels with the term involving the radial derivative of the Laplacian. We are then left with 
\begin{equation}\label{eq:radial_magnetic_field}
    B_r = -r\nabla^2_{ang}\Phi = -\nabla^2_{ang} (r\Phi),
\end{equation}
where 
\begin{equation}
    \nabla^2_{ang} = \frac{1}{r^2 \sin{\theta}}\frac{\partial}{\partial \theta} \left(\sin \theta \frac{\partial}{\partial \theta} \right) + \frac{1}{r^2 \sin^2{\theta}}\frac{\partial^2}{\partial \varphi^2}
\end{equation}
is the angular part of the Laplacian.

\section{Rotating a Vector Field in Spherical Coordinates}\label{app:tilt_vector_field}

In this Appendix, we explain the steps taken to perform a rotation of a vector field in spherical coordinates.

While a spherical geometry is ideal for simulating NS magnetospheres, spherical coordinates can often be cumbersome because of the non constant nature of the unit vectors and the inherent divergence at the origin and at the axis. Tasks such as rotating a vector field are much easier to deal with in Cartesian coordinates. 

The steps that need to be followed to rotate a vector field $\bm{B}$ by an angle $\beta$ around the x-axis are 
\begin{enumerate}
    \item Transform to Cartesian coordinates
    \item Rotate by an angle $-\beta$
    \item Transform back to spherical coordinates
    \item Express the vector field in the rotated coordinates
    \item Transform the vector components to Cartesian
    \item Rotate by an angle $\beta$
    \item Transform the vector components back to spherical 
\end{enumerate}
The spherical to Cartesian transformation is given by the relations

\begin{align}
    x &= \frac{1}{q} \sin {\theta} \cos {\phi} \\
    y &= \frac{1}{q} \sin {\theta} \sin {\phi} \\
    z &= \frac{1}{q} \cos {\theta}.
\end{align}
Rotation by an angle $-\beta$ around the x-axis is straightforward
\begin{equation}
    \begin{pmatrix}
        x' \\
        y' \\
        z'
    \end{pmatrix} 
    = 
    \begin{pmatrix}
        1 & 0 & 0 \\
        0 & \cos {\beta} & \sin {\beta} \\
        0 & -\sin {\beta} & \cos {\beta}
    \end{pmatrix} 
    \begin{pmatrix}
        x \\
        y \\
        z
    \end{pmatrix}.
\end{equation}
Transformation back to spherical is given by the relations
\begin{align}
    q' &= (x'^2 + y'^2 + z'^2) ^{-\frac{1}{2}}\\
    \theta' &= \arccos{(z' q')}\\
    \phi' &= \arctan2 \ (y', x'),
\end{align}
where arctan2 is the 2-argument arctangent, that ensures that the correct and unambiguous value of $\phi$ is returned. By definition, arctan2 returns values in the interval $(-\pi, \pi]$. In order to map $\phi'$ to the desired interval $[0, 2 \pi)$, we add $2 \pi$ to all the negative values. Besides, one can easily see that $q' = q$, which is expected since the radius does not change with the rotation. 

To rotate any scalar field, it is enough to express it in these new, rotated coordinates. However, for vector fields a rotation of the components is also needed. First the vector field $\bm{B}$ is expressed in the rotated coordinates such that $B_q' = B_q (q', \theta', \phi')$, $B_\theta' = B_\theta (q', \theta', \phi')$, $B_\phi' = B_\phi (q', \theta', \phi')$. Then, we map the vector field to the Cartesian unit vectors
\begin{equation}
    \begin{pmatrix}
        B_x' \\
        B_y' \\
        B_z'
    \end{pmatrix} 
    = 
    \begin{pmatrix}
        \sin{\theta'} \cos{\phi'} & \cos{\theta'} \cos{\phi'} & -\sin{\phi'} \\
        \sin{\theta'} \sin{\phi'} & \cos{\theta'} \sin{\phi'} & \cos{\phi'} \\
        \cos{\theta'} & -\sin {\theta'} & 0
    \end{pmatrix} 
    \begin{pmatrix}
        B_q' \\
        B_\theta' \\
        B_\phi'
    \end{pmatrix}.
\end{equation}
and perform a rotation of angle $\beta$
\begin{equation}
    \begin{pmatrix}
        \Tilde{B_x} \\
        \Tilde{B_y} \\
        \Tilde{B_z}
    \end{pmatrix} 
    = 
    \begin{pmatrix}
        1 & 0 & 0 \\
        0 & \cos {\beta} & \sin {\beta} \\
        0 & -\sin {\beta} & \cos {\beta}
    \end{pmatrix} 
    \begin{pmatrix}
        B_x' \\
        B_y' \\
        B_z'
    \end{pmatrix}.
\end{equation}
The last step is to transform the rotated components that have been calculated in the rotated coordinate system back to spherical
\begin{equation}
    \begin{pmatrix}
        \Tilde{B_q} \\
        \Tilde{B_\theta} \\
        \Tilde{B_\phi}
    \end{pmatrix} 
    = 
    \begin{pmatrix}
        \sin{\theta} \cos{\phi} & \sin{\theta} \sin{\phi} & \cos{\theta} \\
        \cos{\theta} \cos{\phi} & \cos{\theta} \sin{\phi} & -\sin{\theta} \\
        -\sin{\phi} & \cos {\phi} & 0
    \end{pmatrix} 
    \begin{pmatrix}
        \Tilde{B_x} \\
        \Tilde{B_y} \\
        \Tilde{B_z}
    \end{pmatrix}.
\end{equation}


\bsp	
\label{lastpage}
\end{document}